\documentclass[a4paper,11pt]{article}

\usepackage{jheppub}
\usepackage[usenames,dvipsnames,svgnames,table]{xcolor}
\usepackage{placeins}
\usepackage{booktabs}
\usepackage{ragged2e}
\usepackage{etoolbox}
\usepackage{multirow}
\usepackage{amssymb}
\usepackage{bbold}
\usepackage{booktabs,colortbl}
\usepackage[normalem]{ulem}
\usepackage{url}
\apptocmd{\thebibliography}{\justifying}{}{}

\makeatletter\g@addto@macro\bfseries{\boldmath}\makeatother

\definecolor{myblue}{rgb}{0.152941176,0.549019608,0.670588235}

\hypersetup{
    pdftoolbar=true,        
    pdfmenubar=true,        
    pdffitwindow=false,     
    pdfstartview={FitH},    
    pdftitle={Machine learning the trilinear and light-quark Yukawa couplings from Higgs pair kinematic shapes},
    pdfauthor={Lina Alasfar, Ramona Groeber, Christophe Grojean, Ayan Paul and Zhuoni Qian},
    pdfkeywords={Future Colliders} {dihiggs} {Machine Learning} {Shapley Values},
    pdfnewwindow=true,
    colorlinks=true,
    linkcolor=myblue,
    citecolor=myblue,
    filecolor=myblue,
    urlcolor=myblue,
    linktocpage=true
}

\def\equationautorefname~#1\null{Eq.\,(#1)\null}
\newcommand{\appendixref}[1]{\hyperref[#1]{appendix~\ref{#1}}}

\newcommand*{\gev}{\text{GeV}}
\newcommand*{\tev}{\text{TeV}}

\newcommand*{\iab}{\ensuremath{\text{ab}^{-1}}}

\newcommand{\inab}{\,{\rm ab}^{-1}}

\newcommand{\bbh}{b\bar bh}
\newcommand{\tth}{t\bar th}

\newcommand{\bbaa}{b\bar b\gamma\gamma}

\newcommand{\hhbox}{hh^{gg\rm F}_{\rm box}}
\newcommand{\hhtri}{hh^{gg\rm F}_{\rm tri}}
\newcommand{\hhint}{hh^{gg\rm F}_{\rm int}}
\newcommand{\qqA}{q\bar q hh}
\newcommand{\uuA}{u\bar u hh}
\newcommand{\ddA}{d\bar d hh}
\newcommand{\QQh}{Q\bar Q h}

\newcommand{\rb}[1]{\rotatebox{90}{#1}}

\title{Machine learning the trilinear and light-quark Yukawa couplings from Higgs pair kinematic shapes}

\author[a]{Lina Alasfar,}
\author[b]{Ramona Gr\"ober,}
\author[a,c]{Christophe Grojean,}
\author[a,c]{Ayan Paul,}
\author[d]{\\and Zhuoni~Qian}

\affiliation[a]{Institut f\"ur Physik, Humboldt-Universit\"at zu Berlin, D-12489 Berlin, Germany}
\affiliation[a]{Dipartimento di Fisica e Astronomia 'G.~Galilei', Universit\`a di Padova and INFN, Sezione di Padova, I-35131 Padova, Italy}
\affiliation[c]{Deutsches Elektronen-Synchrotron DESY, Notkestr. 85, 22607 Hamburg, Germany}
\affiliation[d]{School of Physics, Hangzhou Normal University, Hangzhou, Zhejiang 311121, China}

\emailAdd{lina.alasfar@physik.hu-berlin.de}
\emailAdd{ramona.groeber@pd.infn.it}
\emailAdd{christophe.grojean@desy.de}
\emailAdd{ayan.paul@desy.de}
\emailAdd{zhuoniqian@hznu.edu.cn}

\abstract{
Revealing the Higgs pair production process is the next big challenge in high energy physics. In this work, we explore the use of interpretable machine learning and cooperative game theory for extraction of the trilinear Higgs self-coupling in Higgs pair production. In particular, we show how a topological decomposition of the gluon-gluon fusion Higgs pair production process can be used to simplify the machine learning analysis flow. Furthermore, we extend the analysis to include $q\bar{q}\to hh$ production, which is strongly suppressed in the Standard Model, to extract the trilinear Higgs coupling and to bound large deviations of the light-quark Yukawa couplings from the Standard Model values. The constraints on the rescaling of the trilinear Higgs self-coupling, $\kappa_\lambda$, and the rescaling of light-quark Yukawa couplings, $\kappa_u$ and $\kappa_d$, at HL-LHC (FCC-hh) from single parameter fits are:
\begin{eqnarray}
    \kappa_\lambda&=&[0.53,1.7] \;\;([0.97, 1.03])\nonumber\\
    \kappa_u&=&[-470,430] \;\;([-58,55])\nonumber\\
    \kappa_d&=&[-360,360] \;\;([-26,28])\nonumber
\end{eqnarray}
We show that the simultaneous modification of the Yukawa couplings can dilute the constraints on the trilinear coupling significantly. We perform similar analyses for FCC-hh. We discuss some motivated flavourful new physics scenarios where such an analysis prevails.
}
\begin{document} 
\begin{flushright}
DESY 22-085\\
HU-EP-21/34-RTG
\end{flushright}

\maketitle

\section{Introduction}
\label{sec:Intro}
As the LHC continues to probe the intricacies of the Higgs boson, its couplings to gauge bosons and third-generation fermions have been determined with an accuracy of around 10\% -- 20\%~\cite{ATLAS:2019nkf, CMS:2018uag, Zyla:2020zbs}. Other couplings such as the couplings of the Higgs bosons to the first two generations and the Higgs self-couplings remain more elusive. The high-luminosity LHC (HL-LHC) will contribute significantly to providing limits on the Higgs trilinear self-coupling and couplings of the first generation of quarks to the Higgs. The former can be measured directly at the HL-LHC in Higgs pair production, where it enters via top quark mediated triangle diagrams via gluon fusion~\cite{Djouadi:1999rca, Baur:2003gp, Dolan:2012rv, Baglio:2012np, Abdughani:2020xfo}. This process also contains box diagrams that do not involve the trilinear Higgs self-coupling. Combining several final state channels, with the $b\bar{b}\gamma\gamma$ being the most sensitive, and combining the results of the ATLAS and CMS experiments a sensitivity reach of $0.5< \kappa_{\lambda}=g_{hhh}/g_{hhh}^{SM}<  1.5$ at 68\% CL can be estimated~\cite{DiMicco:2019ngk}. Alternatively, one can use single Higgs production, where the trilinear Higgs self-coupling enters via electroweak loops, leading to comparable limits on the trilinear Higgs self-coupling~\cite{Gorbahn:2016uoy, Degrassi:2016wml, Bizon:2016wgr, Maltoni:2017ims, Haisch:2021hvy} to the Higgs pair production. However, the sensitivity to the Higgs self-coupling from single Higgs  measurements diminishes once other new physics effects modifying also the single Higgs rates are included~\cite{DiVita:2017eyz, Alasfar:2022zyr}.

Measurements of the second generation quark Yukawa couplings are more challenging than those for the third generation. The couplings of the Higgs boson to the charm quarks may be limited to be smaller than a factor of 8.5 times their SM value~\cite{ATLAS:2022ers, CMS:2019hve} from $VH$ production with subsequent decay of the Higgs boson into charm quarks~\cite{Perez:2015lra}\footnote{For the second-generation lepton sector, the analysis of~\cite{ATLAS-CONF-2019-028} constrains the Higgs couplings to muons by a factor of 1.3 with respect to its SM value.}. Other proposals for constraining the charm quark Yukawa coupling include exclusive Higgs boson decays to vector mesons~\cite{Bodwin:2013gca, Alte:2016yuw}, which have been searched for by ATLAS and CMS~\cite{ATLAS:2018xfc, CMS:2022njd}, Higgs + charm production~\cite{Brivio:2015fxa}, Higgs production in association with a jet~\cite{Bishara:2016jga}, as well as the $VVcj$ channel~\cite{Vignaroli:2022fqh}. The strange-quark coupling to the Higgs boson is somewhat more challenging to constrain. However, a future $e^+ e^-$ machine might allow reaching SM sensitivity by making use of strange tagging~\cite{Duarte-Campderros:2018ouv}. The first-generation Yukawa couplings are significantly more difficult to constrain and the aforementioned Higgs pair production can be used to constrain these couplings at future colliders~\cite{Alasfar:2019pmn}.

\begin{figure}[t]
\centering
\includegraphics[width=0.75\textwidth]{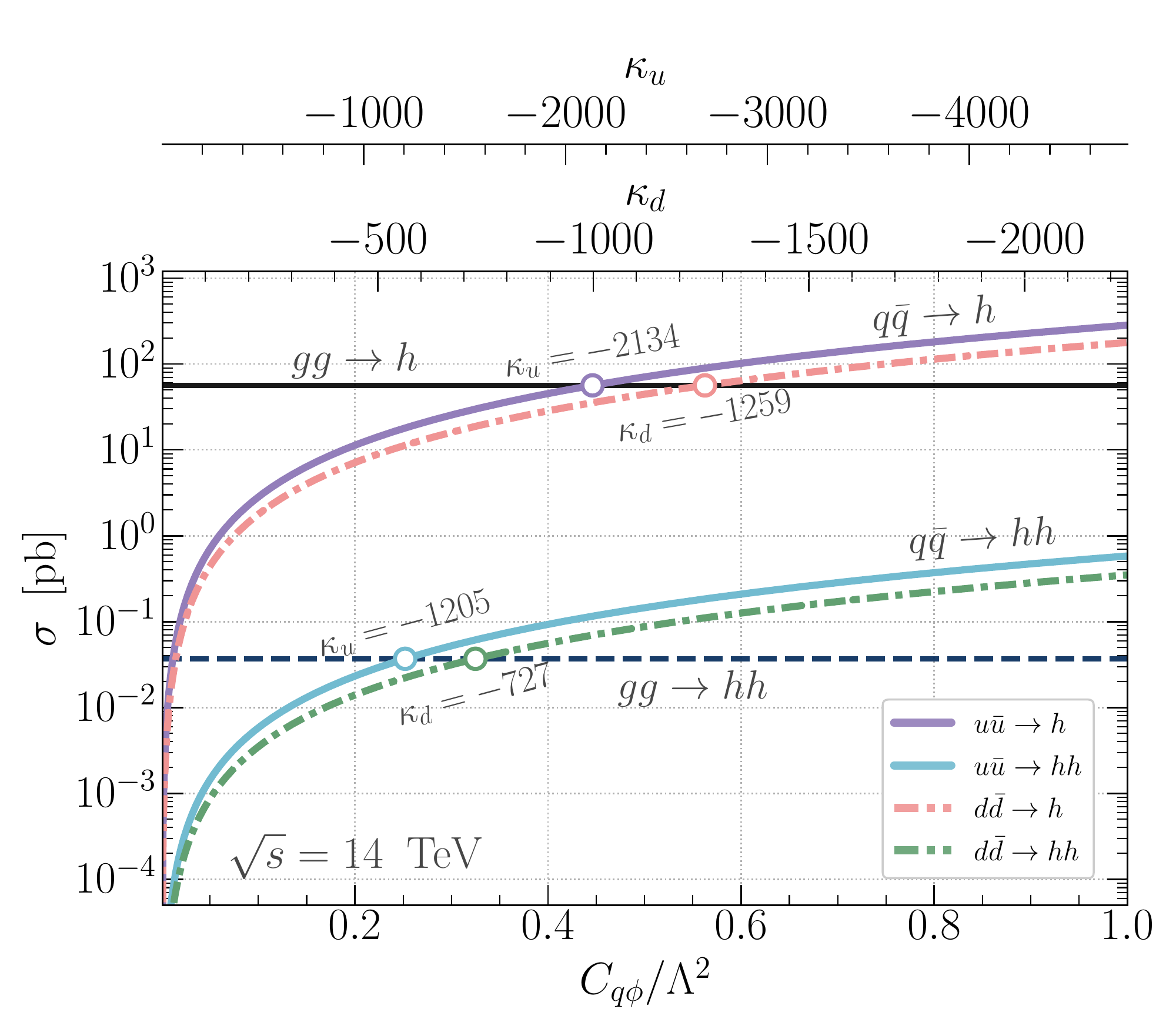}
\caption{\it The production cross-section of single Higgs and di-Higgs at 14 TeV from the quark anti-quark annihilation $q\bar{q}hh$ as a function of the Wilson coefficients $C_{u\phi}/\Lambda^2$ and $C_{d\phi}/\Lambda^2$ versus the SM gluon fusion cross-sections (the horizontal solid lines for $gg \to h$ and $gg \to hh$). One can observe that for values of $\kappa_{u}=-1205\, (-2134)$ and $\kappa_{d}=-727\, (-1205)$ the $q\bar{q}hh$ channels become the dominant di-Higgs (single Higgs) production channels.}
\label{fig:pphhvsh}
\end{figure}

For our analysis, we will adopt  the framework of Standard Model effective field theory (SMEFT) which parameterizes potential new physics effects with all possible higher-dimensional operators under the assumption that the Higgs field has the same transformation rules under the SM symmetries. The leading effects in the Higgs sector are given by dimension six operators. The operators that modify the light-quark Yukawa couplings at the dimension six level are schematically denoted by
\begin{equation}
  \frac{C_{q\phi}}{\Lambda^2} \phi^{\dagger}\phi \bar{Q}_L \phi q_R\,,
\end{equation}
where $Q_L$ are left-handed quark doublets, $q_R$ right-handed quark fields and $\phi$ the Higgs doublet field. This operator also introduces couplings of the light quarks with two or three Higgs bosons. The contribution of the contact interaction of two Higgs bosons with two light quarks in quark annihilation to Higgs pairs is allowed in the SMEFT, and enhances the sensitivity reach on the concerned operator, comparing the Higgs pair production process to the single Higgs production process, as illustrated by \autoref{fig:pphhvsh}. The advantage of the Higgs pair process can be understood primarily from the parton luminosity change when the hard scale goes from the single Higgs mass scale (125 GeV) to the di-Higgs threshold (300 GeV for $q\bar q$ and 400 GeV for $gg$-fusion process, read from the peak of $m_{hh}$ distribution in Ref.~\cite{Alasfar:2019pmn}). The parton luminosity reduction from single Higgs to di-Higgs process is correspondingly 0.44 (0.37) for the $u\bar u (d\bar d)$ channels and 0.09 for the $gg{\rm F}$ channel.\footnote{For 100 TeV, with the parton luminosity estimated at the same typical scale, the corresponding numbers are now 0.97 (0.96) for the $u\bar u (d\bar d)$ channels and 0.90 for the $gg{\rm F}$ channel.} This factor of five differences dictates the cross-section difference. As the $q\bar q$ process is proportional to the square of the coupling, and the crossing point of $C_{q\phi}$ is the square-root of the cross-section ratio between $gg{\rm F}$ and $q\bar q$, the difference between single Higgs and di-Higgs cross-sections is about a factor of $\sqrt{5}\sim 2$. In short, the $q\bar q$-fusion process gains sensitivity at a higher scale mostly through the milder drop of parton luminosity compared to the SM $gg{\rm F}$ processes. For the total cross-section, there is also a ``dilution'' factor from the modified total width of the Higgs for a final state of a specific (non-``light-jet'') decay channel. 

Machine learning techniques are extremely useful in the detection and analyses of the Higgs boson pair production. Boosted decision trees (BDTs) are used both by the ATLAS and CMS collaboration to tag bottom quarks~\cite{DiMicco:2019ngk}. An increased sensitivity to the trilinear Higgs self-coupling can be achieved by employing neural networks or BDTs~\cite{ATLAS:2019vwv,Adhikary:2020fqf,Amacker:2020bmn,Tannenwald:2020mhq,Huang:2022rne}. The present analysis aims to go a step further. While the interpretation of results obtained by machine learning approaches remains notoriously difficult when machine learning is used as a ``black box'', the use of Shapley values, a measure derived from Coalition Game Theory, provides an interpretable analysis framework. We use this framework for the analysis of the $b\bar{b}\gamma\gamma$ final state from Higgs pair production, in  a manner similar to what was previously proposed for the $b\bar{b}h$ process~\cite{Grojean:2020ech,Grojean:2022mef}. This technique is used to extract the trilinear Higgs self-coupling and to probe the light-quark Yukawa couplings from the measurement of kinematics shapes in Higgs pair production significantly improving on the results from the cut-based analysis~\cite{Alasfar:2019pmn}.

The paper is structured as follows: in \autoref{sec:EFT} we show the relevant EFT operators for this analysis and briefly discuss some concrete examples of how large light Yukawa coupling modifications can be realised in models of new physics (NP) making use of the concept of aligned flavour violation (AFV) in order to avoid stringent flavour bounds. In \autoref{sec:Sim} we introduce the details of the simulation of the $pp\to hh \to b\bar b\gamma\gamma$ channel and its backgrounds. In \autoref{sec:kinematics} we discuss the multivariate analysis and the interpretable machine learning approach we adopt. We present prospective results in \autoref{sec:hadronC} for the HL-LHC and the Future Circular Collider (FCC-hh) with a center of mass energy of $\sqrt{s}=100\text{ TeV}$ and compare them to existing and projected bounds. In \autoref{sec:Sum} we summarize our main findings. A discussion on the theoretical uncertainties can be found in \autoref{sec:errors} and a brief overview of prospective bounds on the light-quark Yukawa couplings and Higgs trilinear coupling at future lepton colliders can be found in \autoref{sec:Lep}. Details on the mathematical formulation of Shapley values can be found in \autoref{sec:shapley}.

\section{New Physics from Higgs pair production}
\label{sec:EFT}

The potential deformations of the SM in a model-independent manner can be studied through an EFT description parameterizing NP with higher-dimensional operators suppressed by some large energy scale $\Lambda$. A complete basis for the higher-dimensional operators has been given in Refs.~\cite{Grzadkowski:2010es,Contino:2013kra}. In this work, we are interested in probing the Higgs trilinear and light-quark Yukawa couplings. Starting with the dimension six operators modifying the Higgs self-couplings, we see that they are given by 
\begin{align}
\mathcal{L} \supset &
\frac{C_{\phi\Box}}{\Lambda^2}\,(\phi^\dagger \phi)\Box(\phi^\dagger \phi)+\frac{C_{\phi D}}{\Lambda^2}\,(\phi^\dagger D_\mu \phi)^*(\phi^\dagger D^\mu \phi)+\frac{C_\phi}{\Lambda^2}|\phi^{\dagger} \phi|^3,
\label{eq:EFTop}
\end{align}
where $\phi$ denotes the Higgs-doublet which, in the unitary gauge, can be written as $\phi=1/\sqrt{2}(0,v+h)^T$. It is common to quote the constraints on the Higgs couplings in terms of the rescaling with respect to the SM coupling prediction, typically denoted by $\kappa$:
\begin{equation}
    \kappa = \frac{g_h}{g_h^{\mathrm{SM}}}\,.
\end{equation}
If the NP contributions do not generate new Lorentz structures, there is a possible translation between the Wilson coefficients in the Warsaw basis of the SMEFT, and the $\kappa$ formalism used to denote the rescaling of SM couplings. In particular, taking the rescaling of the trilinear coupling, $\kappa_\lambda$, the translation is given by
\begin{equation}
    \kappa_\lambda = 1-\frac{2 v^4}{m_h^2} \frac{C_\phi}{\Lambda^2}+3 C_{\phi,\mathrm{kin}},
\end{equation}
with
\begin{equation}
  C_{\phi,\mathrm{kin}} = \left( C_{\phi\Box} -\frac{1}{4} C_{\phi D}\right) \frac{v^2}{\Lambda^2},
\end{equation}
and $m_h=125.1 GeV$. The Wilson coefficients $C_{\phi\Box}$ and $C_{\phi D}$ modify all the Higgs couplings and are in parts strongly constrained by electroweak precision observables (e.g.~the $T$ parameter constrains $C_{\phi D}$)~\cite{Ethier:2021bye,Paul:2022dds,DiLuzio:2022xns}. Therefore, we set~$c_{\phi,\mathrm{kin}}=0$ in what follows.

In SMEFT, new flavour structures can be introduced through dimension six operators which contain flavour indices. Focusing on the quark coupling to the Higgs boson we have, in particular,
\begin{align}
\mathcal{L} \supset \frac{\phi^{\dagger}\phi}{\Lambda^2}\left( (C_{u\phi})_{ij} \bar{q}_L^i \tilde{\phi} u_R^j + (C_{d\phi})_{ij} \bar{q}_L^i \phi d_R^j +h.c.\right)\,, \label{eq:lightyukmod}
\end{align}
with $i,j=1\ldots3$. Here $u$ and $d$ refer to the up- and down-type sectors respectively, and not the quarks themselves. The mass matrices of the up- and down-type quarks obtained from the Yukawa and the new SMEFT coupling are
\begin{align}
M^u_{ij} =& \frac{v}{\sqrt{2}} \left( y^u_{ij}-\frac{1}{2} (C_{u\phi})_{ij}\frac{v^2}{\Lambda^2}\right)\,,\nonumber\\
M^d_{ij} =& \frac{v}{\sqrt{2}} \left( y^d_{ij}-\frac{1}{2} (C_{d\phi})_{ij}\frac{v^2}{\Lambda^2}\right)\,. \label{eq:mass}
\end{align}
Due to the modification of the mass matrix, the rotation matrices transforming the quark wavefunction to the mass eigenbasis will be modified with respect to the SM ones. The matrices $C_{q\phi}$ ($q=u,d$) are rotated by a new set of bi-unitary transformations~$\mathcal{V}_{L/R}^{u/d}$ that rotate the quark wavefunctions to the mass eigenbasis. We can write~$C_{q\phi}$ in terms of ~$\tilde{C}_{q\phi}$ which are now in the mass eigenbasis:
\begin{align}
(C_{u\phi})_{ij}&=(\mathcal{V}_{L}^{u})_{li}(\tilde{C}_{u\phi})_{lm} (\mathcal{V}_R^{u})^\dagger_{mj},\nonumber \\
(C_{d\phi})_{ij}&=(\mathcal{V}_{R}^{d})^\dagger_{li} (\tilde{C}_{d\phi})_{lm} (\mathcal{V}_L^{d})_{mj}.
\label{eq_defV}
\end{align}
From this, we see that flavour off-diagonal couplings can be generated in the mass basis that drive flavour-changing neutral currents (FCNCs) which we shall discuss later in this section. The couplings of a single and pair of Higgs bosons to fermions in the mass eigenbasis can be defined as
\begin{equation}
\mathcal{L}\supset g_{h\bar{q}_i q_j}\bar{q}_i q_j h + g_{h\bar{q}_i q_j}\bar{q}_i q_j h^2\,,
\end{equation} 
with
\begin{equation}
g_{h\bar{q}_i q_j} = \frac{m_{q_i}}{v}\delta_{ij}-\frac{v^2}{\Lambda^2} \frac{(\tilde{C}_{q\phi})_{ij}}{\sqrt{2}}\,, \qquad g_{h h\bar{q}_i q_j} = -\frac{3}{2\sqrt{2}}\frac{v}{\Lambda^2}(\tilde{C}_{q\phi})_{ij}\,. 
\label{eq:couplingsEFT}
\end{equation}
We have assumed that the matrices $\tilde{C}_{q\phi}$ are real, otherwise strong constraints from electric dipole moments apply~\cite{Chien:2015xha, Brod:2018lbf}. A similar relation exists for the rescalings of the quark Yukawa couplings~$\kappa_q$, for the diagonal elements of~$\tilde{C}_{q\phi}$\footnote{For normalisation purposes, we choose to set $\Lambda = 1\,\mathrm{TeV}$ throughout the remainder of this paper. We stress that it does not imply that the scale of new physics is at 1 TeV, but it fixes the normalisation of the Wilson coefficients.}
\begin{equation}
  \kappa_{q_i} = 1- \frac{v^3}{\sqrt{2}m_q}\frac{(\tilde{C}_{q\phi})_{ii}}{\Lambda^2},
\end{equation}
with $m_u=2.2 MeV$ and $m_d=4.7 MeV$. The $hh q\bar{q}$ coupling, though being linearly related to the quark Yukawa coupling $h q\bar{q}$, is not a rescaling of any SM Higgs coupling. With this in mind, one can remain strictly within an EFT, like SMEFT, where the $SU(2)\times U(1)$ symmetry is linearly realized and links the rescaling of the quark Yukawa, $\kappa_q$, to the~$hh q\bar{q}$ coupling through
\begin{equation}
  g_{hhq\bar{q}}^{\mathrm{linear-EFT}} = \frac{3}{2}\frac{\kappa_q-1}{v} \, g_{h q\bar{q}}^{\mathrm{SM}}.
\end{equation}
This relation will no longer hold once a non-linear EFT is used. Hence, some caution has to be exercised when the $\kappa$-formalism is used in multi-Higgs studies. For clarity we will denote $\tilde{C}_i$ as $C_i$, i.e., assuming $C_i$ to be in the mass basis, in the rest of the article.

As we have pointed out before, the construction we discussed can lead to contributions from FCNCs which are strongly constrained from low-energy measurements of flavour observables. The bounds are of order $|(C_{u\phi,d\phi})_{12}| \lesssim 10^{-5}\Lambda^2/v^2$ and $|(C_{u\phi,d\phi})_{13}| \lesssim 10^{-4} \Lambda^2/v^2$ and stem from $\Delta F=2$ transitions~\cite{Blankenburg:2012ex, Harnik:2012pb}. Given that FCNCs need to be suppressed, an effective way of realizing this suppression is by imposing minimal flavour violation (MFV)~\cite{DAmbrosio:2002vsn}. The assumption of MFV introduces a strong hierarchy amongst the Higgs couplings to quarks, due to the proportionality of the Wilson coefficients to the Yukawa couplings. Since we want to explore rather large modifications of the light-quark Yukawa couplings, very low values of $\Lambda$, the NP scale, or large Wilson coefficients in MFV models or both need to be assumed which render the validity of the EFT questionable. Furthermore, this would potentially generate conflicts with measurements of the third generation couplings to the Higgs boson. To escape the strong hierarchy in the Yukawa couplings imposed by assuming MFV, a less restrictive assumption about the flavour structure can be made which is called flavour alignment~\cite{Pich:2009sp,Pich:2010ic}. In a more generalized form of this ansatz, it is assumed that the new physics contributions to the Yukawa matrices are aligned in flavour space and can be simultaneously diagonalized to remove the possibility of having flavour non-diagonal couplings that can generate FCNCs at the tree level~\cite{Ferreira:2010xe,Jung:2010ik,Botella:2015yfa}.

\begin{figure}
    \centering
    \includegraphics[width=0.375\textwidth]{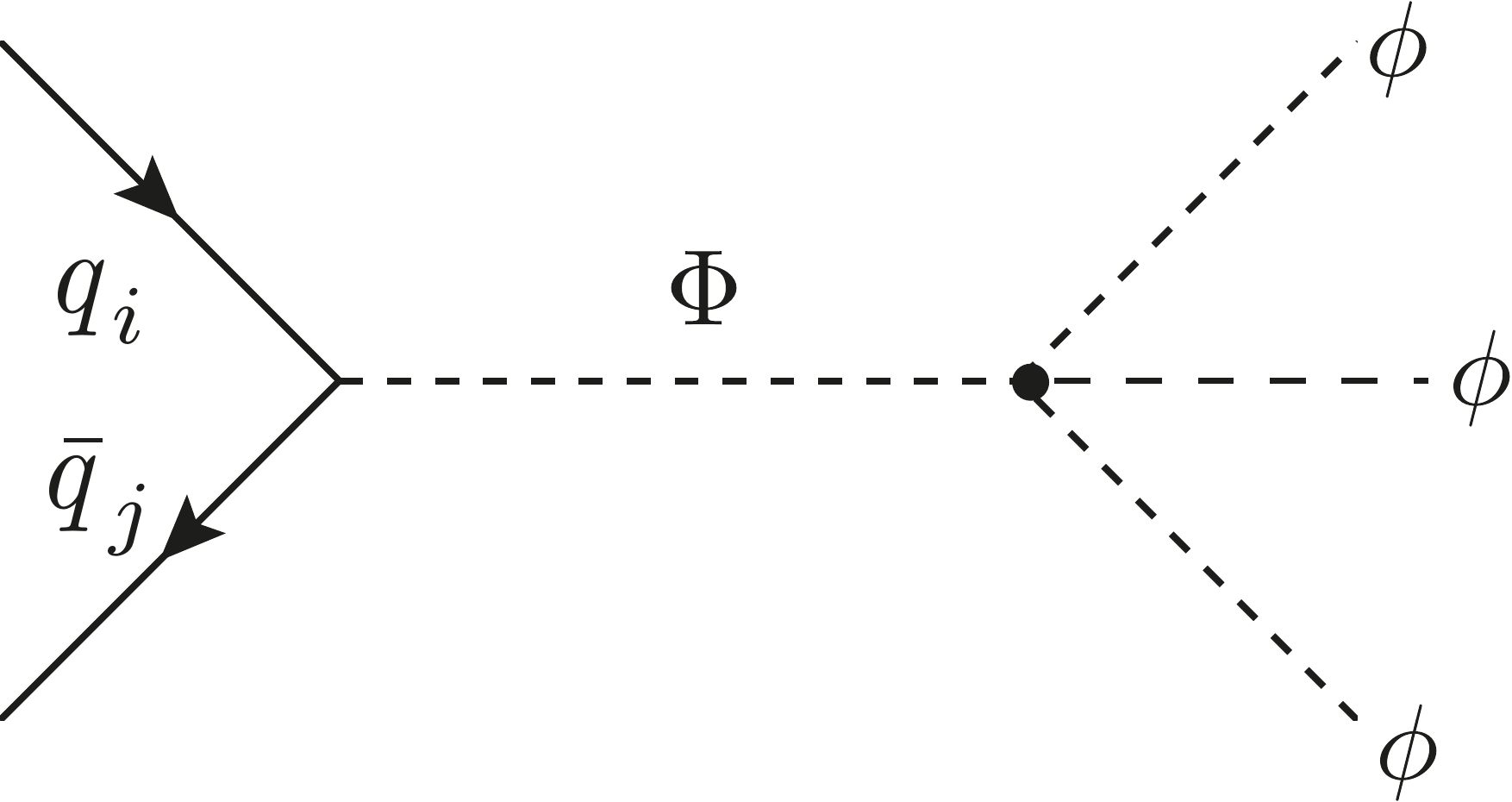}
    \hspace{0.5 cm}
    \includegraphics[width=0.25\textwidth]{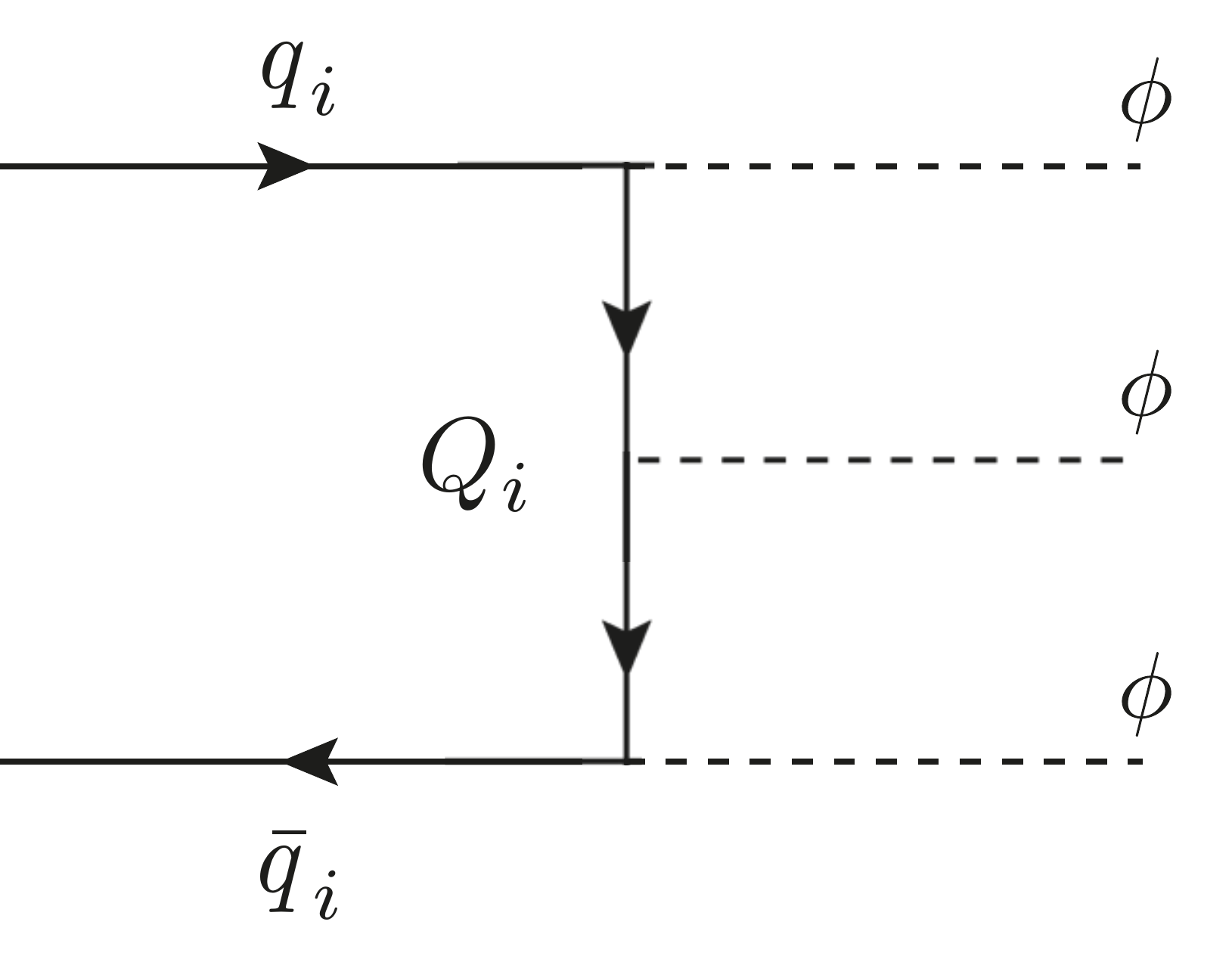}
    \caption{{\it The SMEFT operator $\mathcal{O}_{q\phi}$ can be generated either by a new scalar $\Phi$ (left) or a new set of VLQs $Q_i$ (right). When these fields are integrated out, they yield the aforementioned operator in the unbroken phase.}}
    \label{fig:unbroken-cqphi}
\end{figure}

There exist a handful of potential UV-complete models in which both light-quark Yukawa couplings and the Higgs trilinear couplings are simultaneously enhanced. In general, the operator of \autoref{eq:lightyukmod} can be generated at tree-level by vector-like quarks~(VLQs) or new scalars~\autoref{fig:unbroken-cqphi}. For example, a model proposed in Ref.~\cite{Bar-Shalom:2018rjs} based on VLQs with the assumption of AFV where the mechanism of suppressing FCNCs is linked to flavour textures imposed by discrete symmetries. The original assumption of this model is excluded as the authors assumed that all the light quark-Higgs couplings to be similar in magnitude to the bottom Yukawa. One could still get a significant enhancement of the light-quark Yukawa couplings from VLQ masses of $\sim 2$ TeV, which is well above the bounds set by current direct searches excluding VLQ of masses~$M_{VLQ} < 1.6$ TeV~\cite{Unal:2777832,CMS:2019eqb} for the purely hadronic final state, and $M_{VLQ}< 1.2$ TeV for final states with leptons and jets~\cite{CMS:2018wpl}. Values of $M_{VLQ}\sim 2$ TeV would also be in accordance with electroweak precision observables. In addition, the trilinear Higgs coupling could be modified by the inclusion of an additional scalar singlet as proposed in Ref.~\cite{DiLuzio:2017tfn, Falkowski:2019tft, Chang:2019vez}.

Another concrete example of a model with enhanced light-quark Yukawa couplings is a two-Higgs-doublet model (2HDM) model proposed in Refs.~\cite{Egana-Ugrinovic:2019dqu,Egana-Ugrinovic:2021uew}. This model shows a sub-class of AFV, known as spontaneous flavour violation~\cite{Egana-Ugrinovic:2018znw}. Enhancements of the light-quark Yukawa couplings stem from the couplings of the second Higgs doublet to the quarks, $K_{u/d}$, which are made diagonal in the flavour space. The model can have either the up-type or the down-type couplings enhanced with respect to the SM values, while the couplings of the other type remain proportional to the SM ones in order to obtain the correct CKM matrix. The addition of the second doublet modifies the Higgs potential, and consequently, the Higgs self-coupling will be modified as well. Like any other 2HDM, the parameter space is rather large. Hence, bounds on the model depend on the part of the parameter space that is under consideration. For a small mass of the ``heavy'' Higgs $H$ and large Yukawa coupling, $K_d$, flavour bounds dominate, while for a larger $m_H$, dijet searches~\cite{Aaboud:2019zxd,Aad:2019hjw,Sirunyan:2019vgj} dominate due to the decay $ H \to d \bar d$.  Instead, the decay $H \to hh$  becomes dominant from smaller values of $K_d$ and larger $H$ mass, rendering in the regime $m_H < 2$ TeV resonant di-Higgs searches~\cite{Sirunyan:2018ayu, Aad:2019uzh} as the potential discovery channel of such a scenario. Furthermore, in this region of the parameter space constraints can be derived from $Zh$~\cite{ATLAS-CONF-2020-043} and $ZZ$~\cite{ATLAS:2020tlo, Sirunyan:2018qlb} searches.  Lastly, for~$ m_H > 2$ TeV, the non-resonant Higgs pair production will become the dominant bound on the enhancement of the light-quark Yukawa couplings.
 \section{Events simulation for HL-LHC and FCC-hh}
\label{sec:Sim}

We consider the final state $b \bar{b} \gamma \gamma$ as this channel has the highest potential for Higgs pair searches~\cite{Cepeda:2019klc}. One has the ``clean'' decay $h \to \gamma \gamma$ of one Higgs boson, with the other Higgs boson decaying to a $b\bar b$ quark pair with a large branching ratio~$\sim 58\%$, where $b$-tagging capabilities for ATLAS and CMS are continuously improving. We consider the $b\bar bh$, $t\bar th$, $b\bar b \gamma\gamma$ processes as the main sources of background for the $hh$ signal. The details of the simulation for the $b\bar  b h$ process  can be found in Ref.~\cite{Grojean:2020ech}. The events are generated at leading order (LO) and then scaled to NLO by $K$-factors, defined as the ratio of higher order cross section over its LO counterpart. The $K$-factors are taken from the corresponding literature for $t\bar{t}h$~\cite{Beenakker:2001rj}, $b\bar b \gamma\gamma$~\cite{Fah:2017wlf}, $Zh$~\cite{Campanario:2014lza} and the remaining part of the $b\bar bh$ processes from~\cite{Dawson:2005vi}. The Higgs particles are further decayed to $\gamma\gamma$ following the Higgs cross-section working group recommendations~\cite{LHCHiggsCrossSectionWorkingGroup:2016ypw}. The parton-level results are generated using \texttt{MadGraph\_aMC@NLO}~\cite{Alwall:2014hca}, showered using \texttt{Pythia 8.3}~\cite{Sjostrand:2014zea}, and a subsequent detector simulation is done using \texttt{Delphes 3}~\cite{deFavereau:2013fsa}. To be inclusive and to explore the capabilities and importance of the full detector coverage, no generator-level cuts are applied on these processes except for the $b\bar b \gamma\gamma$ QCD-QED background processes to avoid divergences. These minimal generator-level cuts are
\begin{equation}
    \begin{aligned}
    & Xp_T^b>20\,\gev, \\
    \textrm{generator level cuts:}\qquad& \eta_\gamma<4.2,~ \Delta R_{b\gamma}>0.2, \\
    & 100\, \gev< m_{\gamma\gamma} \, < 150\,\gev.
    \end{aligned}
\end{equation}
Here $Xp_T$ implies a minimum $p_T$ cut for at least one $b$-parton. After the showering and detector simulation, further basic selection cuts are applied to select events with
\begin{equation}
    \begin{aligned}
    & n_{\mathrm{eff}}^{bjet}\geq 1,~n_{eff}^{\gamma jet} \geq 2,\\
    \textrm{basic cuts:}\qquad& p_T^{bjet}>30\,\gev,~p_T^{\gamma jet}>5\,\gev,\\
    & \eta_{bjet,\gamma jet}<4,~ 110\, \gev < m_{\gamma\gamma} < 140\,\gev,
    \end{aligned}
    \label{eqn:bcuts}
\end{equation}
with $n_{\rm eff}^{b/\gamma jet}$ representing the number of $b/\gamma$-jets that pass the basic selection. The cross-section, $K$-factors, number of events with 6 ab$^{-1}$ luminosity at 14 TeV are given in \autoref{tab:xsec14}. 

\begin{figure}[t]
    \centering
    \includegraphics[width=0.25\textwidth]{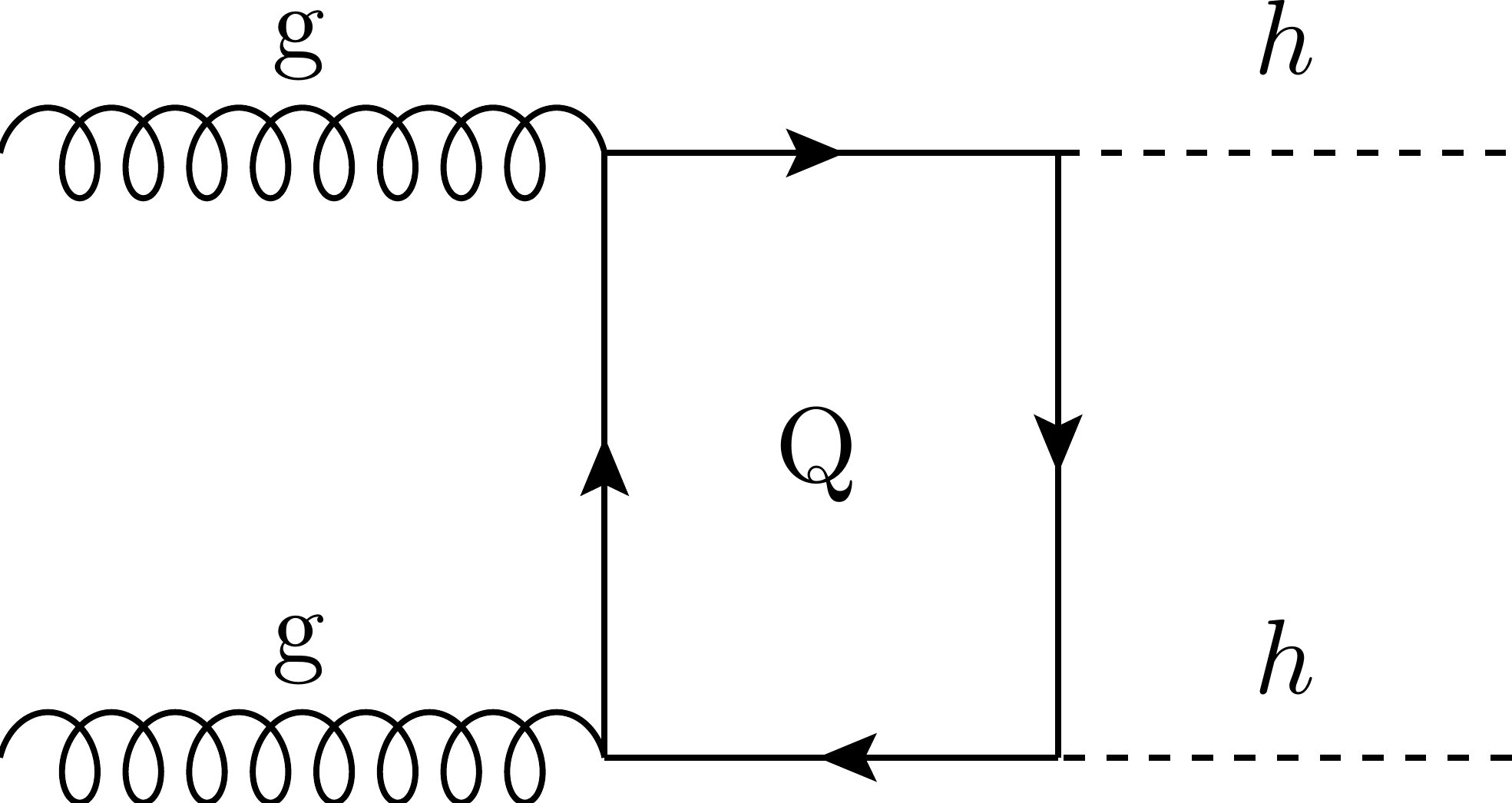} 	\hspace*{0.5 cm}
    \includegraphics[width=0.29\textwidth]{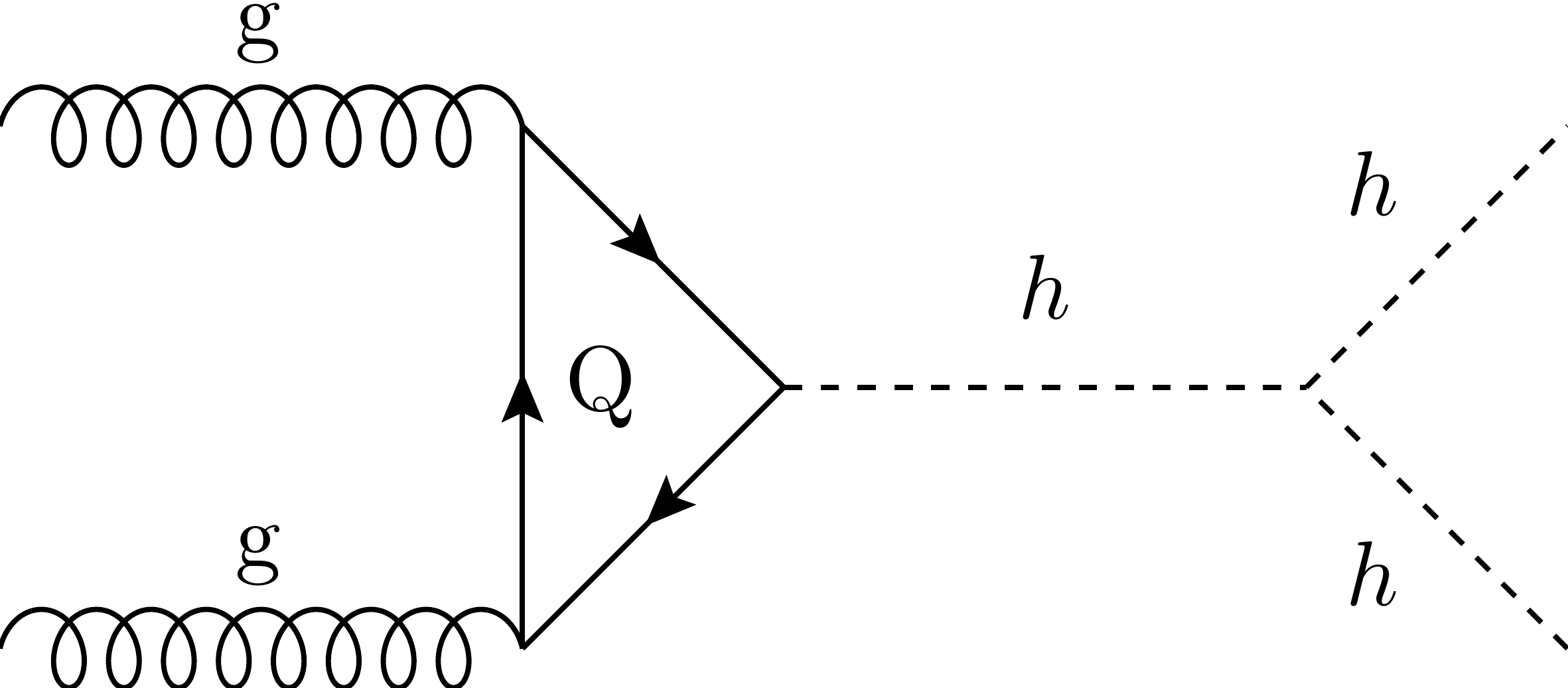}
    \caption{\it The cross-section of the $gg$F channel can be decomposed into three subprocesses based on their dependence on the trilinear Higgs self-coupling, $\lambda$. The triangle topology depends on $\lambda^2$, the box topology does not depend on $\lambda$ and the interference amongst the latter two is linear in $\lambda$.}
    \label{fig:ggfsub}
    \vspace{0.4cm}
	\includegraphics[width=0.125\textwidth]{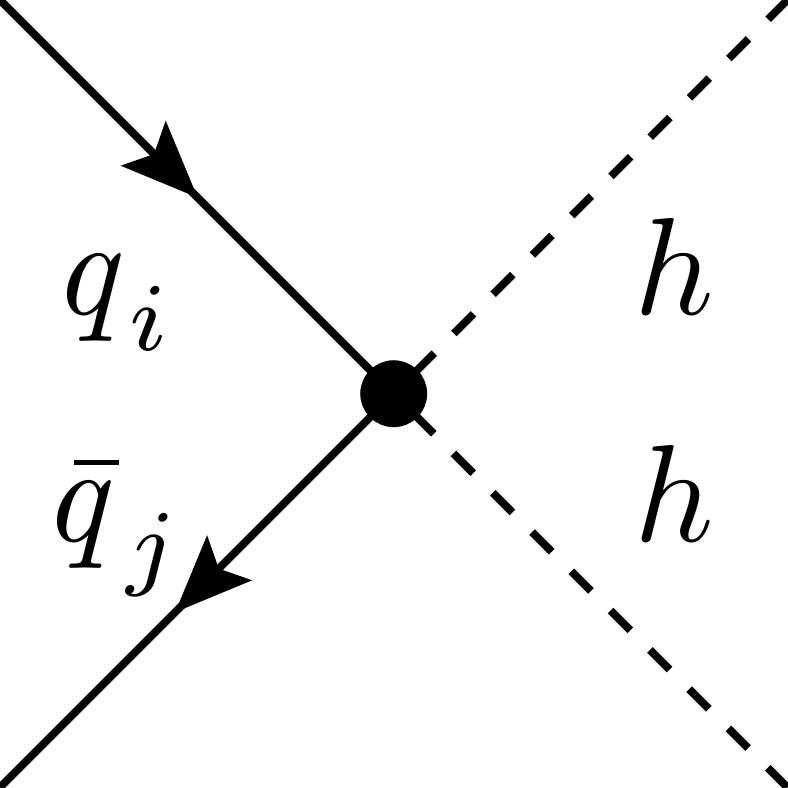}
	\hspace*{0.5 cm}
	\includegraphics[width=0.25\textwidth]{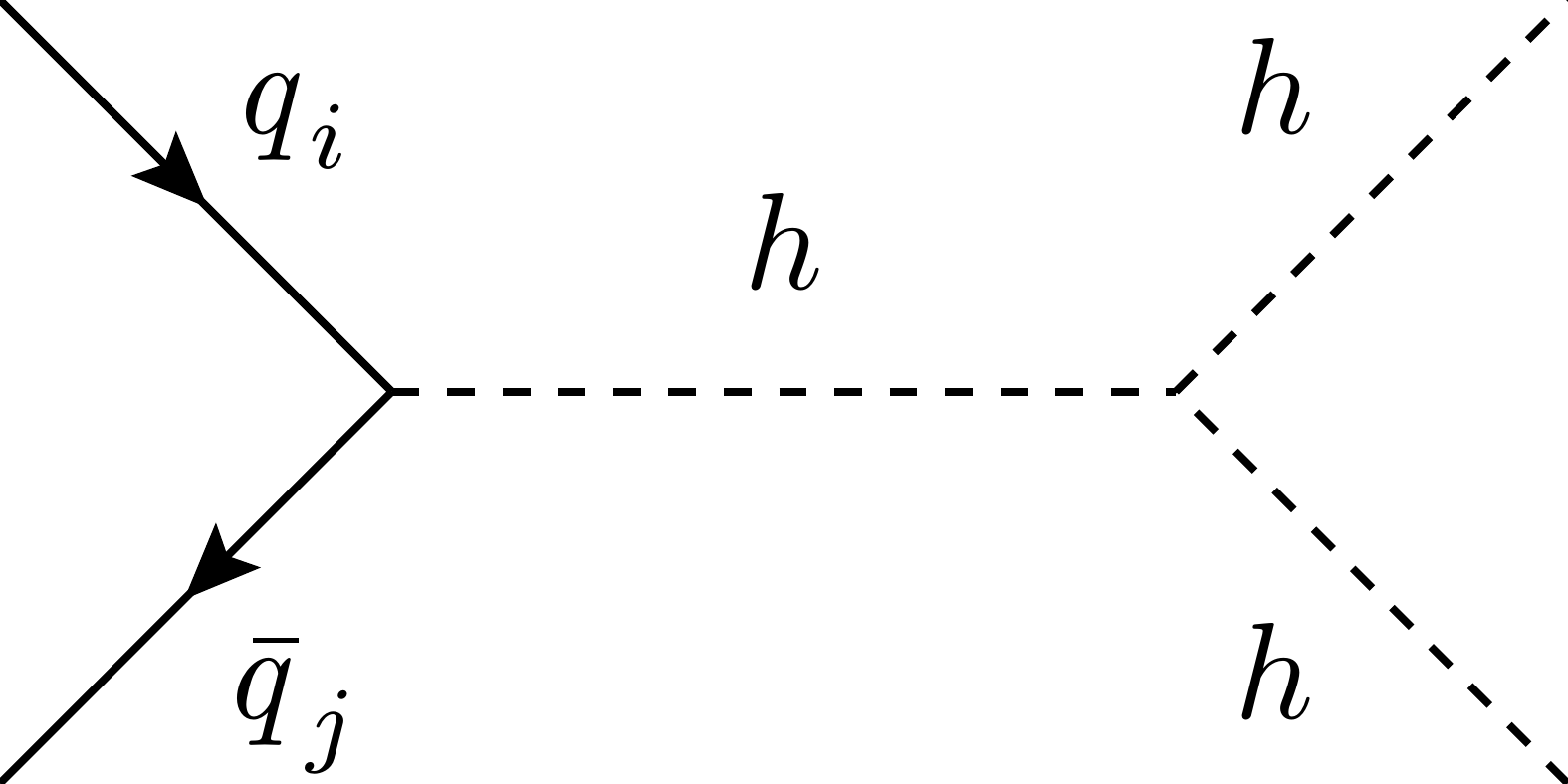}
	\caption{\it The dominant Feynman diagrams for the quark anti-quark annihilation~($q\bar qhh$)  production of Higgs pair, via the SMEFT operator $ \mathcal O_{q\phi}$.}
	\label{fig:qqa}
\end{figure}

\begin{table}[t]
    \centering
    \begin{tabular}{cccc}
    \specialrule{.8pt}{0pt}{0pt}
    Channel	        &LO $\sigma$ [fb]	&NLO $K$-fact	&6$\inab$ [\#evt @ NLO]   \\ 
    \specialrule{.8pt}{0pt}{0pt}
    $y_b^2$	        &0.0648	            &1.5	    &583                \\
    $y_by_t$        &-0.00829	        &1.9        &-95                \\
    $y_t^2$	        &0.123	            &2.5	    &1,840              \\
    $Zh$	        &0.0827	            &1.3	    &645                \\
    $\sum\bbh$	    &0.262	            &-	        &2,970              \\
    \midrule
    $t\bar th$	    &1.156	            &1.2	    &6,938              \\
    \midrule
    $\bbaa$	        &12.9	            &1.5	    &116,000            \\
    \specialrule{.8pt}{0pt}{2pt}
    \end{tabular}
    \caption{\it SM cross-section for the main background processes at 14\,$\tev$ with 6\,$\iab$ data at the HL-LHC, and the number of events after the basic cuts as defined in \autoref{eqn:bcuts}. For $\bbh$ production, the Higgs boson is decayed to a pair of photons. The total production cross-section of Higgs associated with a $b\bar{b}$ pair is denoted by $\sum\bbh$ and is the sum of the top four channels.
    }
    \label{tab:xsec14}
\end{table}

\begin{table}[t]
	\centering
\begin{tabular}{ccccc}
\toprule
 Channel	        &LO $\sigma $ [fb]	&$K$-fact.	&6$\inab$ [\#evt @ order]   \\
	\midrule
	$\hhtri$	        &  $7.3 \cdot10^{-3}$    & 2.28  &  96 (NNLO)  \\ 
	$\hhbox$            & $54 \cdot10^{-3}$    & 1.98 & 680 (NNLO)  \\ 
	$\hhint$            &$-36 \cdot10^{-3}$    & 2.15 &-460 (NNLO)  \\ 
    \midrule
    $\uuA$ $(C_{u\phi}=0.1)$ &  $5.6\cdot10^{-3}$    & 1.30 &  43 (NLO)  \\ 
	$\ddA$ $(C_{d\phi}=0.1)$ &  $3.6\cdot10^{-3}$    & 1.29&  28 (NLO)  \\ 
	\bottomrule
\end{tabular}
	\caption{\it  The LO SM cross-section for di-Higgs production at the HL-LHC for 6 $\inab$ of data multiplied by the $hh \to b \bar b \gamma \gamma$ SM branching ratio, $K$-factors (taken from~\cite{deFlorian:2021azd} for the gluon channels and~\cite{Alasfar:2019pmn} for the quark channels) and the number of events after the basic cuts for the gluon fusion~($gg$F) and quark annihilation~($\qqA$) at $\sqrt{s}=$ 14 TeV. }
	\label{tab:kfact}
\end{table}

The simulation of the $hh$ signal is separated into two main channels. The first is the gluon-fusion~($gg$F) channel which is the dominant channel in the SM and can be further decomposed into three subprocesses based on their dependence on the Higgs trilinear self-interaction, $\lambda$, as seen in \autoref{fig:ggfsub}. Amongst these subprocesses, the first is the amplitude squared of the contribution from the triangle diagram which is proportional to $\lambda^2$. The second is the squared amplitude of the contribution from the box diagram that does not depend on the trilinear coupling. The third is the contribution from the interference between the triangle and box diagrams, which is proportional to  $\lambda$. Using this separation allows us to remove the dependence of the total $K$-factor for $hh$ production on rescaling of the trilinear Higgs coupling~\cite{Heinrich:2019bkc}. The individual $K$-factors for each of the subprocesses are independent of the rescaling of the trilinear Higgs coupling making our analysis computationally much simpler. The $gg$F process is generated using the \texttt{HH} production program implemented in \texttt{POWHEG}~\cite{Heinrich:2017kxx,Heinrich:2019bkc,Heinrich:2020ckp}, which has been modified to separate the individual contributions from the three diagrams. The cross-section for these individual contributions and the corresponding $K$-factors can be found in \autoref{tab:kfact} as derived from  Ref.~\cite{Buchalla:2018yce}. 

The other main process, the quark anti-quark annihilation~($q\bar qhh$), is strongly suppressed in the SM for first-generation quarks since the SM Yukawa couplings are proportional to the mass of the considered quark flavour. However, since this channel is a tree-level process (cf.~\autoref{fig:qqa}), with sufficiently large enhancement factors of the light-quark Yukawa coupling, it becomes dominant as shown in \autoref{fig:pphhvsh}. The $q\bar qhh$ cross-section scales like $C_{q\phi}^2/\Lambda^4$ while the $gg$F production cross-section remains almost unchanged. Therefore, for constraining enhancements of the light-quark Yukawa, we consider this channel as the signal and the $gg$F channel as part of the background. We note that given the fact that the SM contribution is zero if the light quark masses are neglected, the leading SMEFT contributions to this process are (dimension six)$^2$. This does not invalidate the EFT analysis, since the SM$\times$dimension eight contribution again is vanishing. 

The $q\bar qhh$ processes are generated with~\texttt{MadGraph\_aMC@NLO} with a UFO model created with \texttt{FeynRules}~\cite{Alloul:2013bka}. Samples for both up- and down-quark initiated $q\bar qhh$ processes are generated. For all the $hh$ signals, the samples are generated at LO and later scaled by the NLO $K$-factors given in \autoref{tab:kfact}. The $K$-factors are adapted from  Refs.~\cite{Dicus:1998hs,Balazs:1998sb, Harlander:2003ai} as described in~\cite{Alasfar:2019pmn} for the $q\bar qhh$ channel. Moreover, the two Higgs bosons are decayed to $b\bar b$ and $\gamma\gamma$ respectively, with \texttt{Pythia 8.3} and then showered. The same detector simulation and basic cuts as for the background processes are then performed. In addition, the same sets of parton distribution function sets (\texttt{NNPDF31\_nlo\_as\_0118\_nf\_4}) are used for the signal and the background, implemented via \texttt{LHAPDF}~\cite{Buckley:2014ana}. The calculation of the Higgs full width and branching ratios is done using a modified version of \texttt{Hdecay}~\cite{Djouadi:1997yw,Djouadi:2018xqq} to include the effects of SMEFT operators $\mathcal{O}_{q\phi}$  and taken into account as a rescaling factor for the total cross-section of the $hh$ signal and the relevant background processes. It should be noted, that in both di-Higgs production and decay calculation, the light quark masses are set to zero. However, when converting between SMEFT and $\kappa$-formalism, the $\overline{\mathrm{MS}}$ quark masses at $\mu_R= 2$ GeV are used, in accordance with the PDG.

For FCC-hh, almost everything is done similarly after setting the energy to 100 TeV and the luminosity to 30 ab$^{-1}$. Since we do not have all $K$-factors available at a collider energy of 100 TeV we rescaled the LO samples by the same $K$-factors as for HL-LHC. We note that we explicitly checked that at least within the SM, for Higgs pair production via gluon fusion the difference is of $\mathcal{O}(1\%)$~\cite{Maltoni:2014eza} and hence a good approximation.

\section{Exploring higher dimensional kinematic distributions}
\label{sec:kinematics}

After detector simulation and jet definition, we have a final state of two photon jets and at least one identified $b$-jet, where the two photons reconstruct back to a real scalar Higgs mass for all the $\bbh$ channels. We first define and evaluate a comprehensive set of kinematic observables as follows:
\begin{itemize}
\setlength{\itemsep}{0pt}
    \item $p_T^{b_1}$, $p_T^{b_2}$, $p_T^{\gamma_1}$, $p_T^{\gamma\gamma}$, 
    \item $\eta_{b_{j1}}$, $\eta_{b_{j2}}$, $\eta_{\gamma_1}$, $\eta_{\gamma\gamma}$,
    \item $n_{bjet}$, $n_{jet}$, $\Delta R_{\rm min}^{b\gamma}$, $\Delta \varphi_{\rm min}^{bb}$, 
    \item $m_{\gamma\gamma}$, $m_{bb}$, $m_{b_{1} h}$, $m_{b\bar b h}$, $H_T$.
\end{itemize}
Here $p_T^{{b/\gamma}_{1,2}}$ and $\eta^{{b/\gamma}_{1,2}}$ are the transverse momentum and rapidity of the tagged leading and sub-leading $b/\gamma$-jets (in our definition the subleading $b$-jet could be a null four-vector since we require one $b$-jet inclusive), $n_{bj}$ is the number of tagged and passed $b$-jets. The variables $\Delta R_{\rm min}^{b\gamma}$ and $\Delta \varphi_{\rm min}^{bb}$ are the minimum $R$-distance and $\varphi$-angle between a tagged $b$-jet and a photon jet. The remaining variables are the invariant masses and $H_T$ is the scalar sum of the transverse mass of the system. We shall show in what follows that it is not necessary to be very selective about the kinematic variables one chooses to use in the analysis. What is necessary is that all possibly useful kinematic variables are included. As can be seen from the list above, some of the variables seem to be interdependent and, probably, highly correlated. The beauty of using interpretable machine learning is that a hierarchy of importance for the variables will be built during the analysis using an over-complete basis of collider observables from which the most important ones can be chosen to further enhance the analysis.

\subsection{Interpretable machine learning}
\label{sec:BDT}

Rule-based machine learning algorithms have for long been used as the gold standard for a signal to background discrimination in a wide variety of particle physics analyses. They are known to outperform neural networks in terms of simplicity of implementation, computational resources required and accuracy in modelling the underlying distributions.\footnote{Nevertheless, we tested a deep neural network built with Tensorflow~\cite{tensorflow2015-whitepaper} and found no improvement in the classification accuracy.} In addition, rule-based algorithms, such as decision trees, are more transparent as far as separating the signal from the background is concerned. The first tree of the BDT always starts off with the most important kinematic cut which can be seen from the structure of the first tree making the process more interpretable. Placing emphasis on interpretability in multivariate analyses~\cite{Grojean:2022mef}, we chose to work with Boosted Decision Trees (BDT). However, the interpretability of a machine learning algorithm requires more than just a choice of an interpretable model. The conditions are:

\begin{itemize}
\itemsep0em
    \item A variable set that is easily interpretable in terms of the dynamics being studied.
    \item A machine learning algorithm that is more transparent and not a complete black box.
    \item A method for interpreting the model and attribute variable importance to understand how the algorithm models the underlying distributions.
\end{itemize}
Choosing to work with BDTs just satisfies the second condition. For training the BDTs we use XGBoost~\cite{10.1145/2939672.2939785}, a publicly available scalable end-to-end boosting system for decision trees. We follow the normal procedures for training and testing the BDT with simulated data. To satisfy the first condition we chose to work with high-level kinematic variables that are representative of the process instead of working with four-vectors. The disadvantage of working with kinematic variables is that a complete set cannot be defined for a particular process unlike the four-vectors associated with the process. So, in principle, a large number of kinematic variables can be formulated and used in a multivariate analysis. While the number is, in general, not too large for any implementation of BDTs, having a large set of variables clouds the understanding of which ones are important for orchestrating the separation of the signal from the background. This is where the third condition listed above is important. Variable importance attribution is a way to ``short-list'' only those variables that play an important role in the predictive power of the classification (or regression) problem. There are several measures of variable importance used in machine learning like Gini or permutation-based measures~\cite{Breiman2001,JMLR:v20:18-760}, local explanations with surrogate models~\cite{10.1145/2939672.2939778} etc., to name a few. However, these suffer from inconsistencies or fail to provide a global explanation of the model~\cite{NIPS2017_7062}. 

To build a mathematically consistent procedure for variable importance attribution, we use Shapley values~\cite{shapley1951notes} from Coalition Game Theory. Formulated by Shapley in the mid-20$^{th}$ century, Shapley values is a formulation of an axiomatic prescription for fairly distributing the payoff of a game amongst the players in a $n$-player cooperative game. When applied to machine learning, Shapley values tell us how important the presence of a variable is in determining a certain category (like signal or background) when compared to its absence from the multivariate problem being addressed. The process naturally and mathematically lends itself to studying the correlations between different variables since all possible combinations of variables can be taken to check the outcome. A more detailed discussion of the application of Shapley values to signal vs. background classification problems for particle physics can be found in Refs.~\cite{Grojean:2020ech,Alvestad:2021sje,Cornell:2021gut,Grojean:2022mef}. In this work, we follow the same basic procedure as discussed in Ref.~\cite{Grojean:2020ech}. The importance of a variable in determining the outcome of a classification will be quantified by the mean of the absolute Shapley value, $\overline{|S_v|}$, larger values signifying greater importance. We will use the SHAP (SHapley Additive exPlanations)~\cite{NIPS2017_7062} package implemented in python based on Shapley values calculated exactly using tree-explainers~\cite{2018arXiv180203888L,Lundberg:2020vt}.

To provide an intuition of what Shapley values imply let us look at some edge cases. If there are two variables in the problem that are fully correlated then they are considered ``equally good players'' and their Shapley values, axiomatically, will be exactly equal. If there is a kinematic variable that does not contribute to the outcome at all, i.e, if one varies the variable but the outcome remains constant, the Shapley value of that variable will remain exactly zero. Hence, Shapley values encapsulate the correlations, even the higher order ones, between the input variables themselves and also the input variables and the outcome. More details on Shapley values can be found in \autoref{sec:shapley}.

\section{The di-Higgs channel at future hadron colliders}
\label{sec:hadronC}

We would like to study the bounds on three specific couplings in this work. The first one being the Higgs trilinear coupling quantified by $C_\phi$ defined in \autoref{eq:EFTop} and the other two being the deformation of the first-generation SM Yukawa coupling to the up and down quark defined as $C_{u\phi}$ and $C_{d\phi}$ in \autoref{eq:couplingsEFT} with $i=j=1$. We will not consider modifications of the second generation of quarks as their effects in di-Higgs production would be suppressed by the small parton distribution functions. 

In the BDT analysis, we combine the $\bbh ~ (h\to\gamma\gamma)$ and $\tth ~(h\to\gamma\gamma)$ channels into one category calling it $Q\bar Q h$ while the other (continuum) background channel, $\bbaa$, is treated as a separate category. We do not combine all the background channels since the $\bbaa$ channel is, by far, the dominant background and combining the $Q\bar Q h$ channels to it results in an inability of the BDTs to learn the shapes of the $Q\bar Q h$ channels which are comparable in size to the signal. Hence, leaving them separate allows for better classification of the signal. Moreover, this also helps with the interpretability of the classifications in terms of the Shapley values of the kinematic variables. For the analysis involving $C_\phi$, we simulate three separate categories for the triangle, box and interference terms of the $gg$F $hh$ production which we refer to as $\hhtri$, $\hhbox$ and $\hhint$, respectively. The $\qqA$ channels include two other categories, one each for probing the Wilson coefficients $C_{u\phi}$ and $C_{d\phi}$, respectively. Note that all the channels with a Higgs boson as an intermediate state are all sensitive to $C_{u\phi}$ and $C_{d\phi}$ through their modification to the total Higgs width and the $h\to\gamma\gamma$ decay~\cite{Alasfar:2019pmn}, which are all taken into account. In what follows, we refer to the two $\qqA$ channels as $\uuA$ and $\ddA$ explicitly.

As we progress through the analysis we study the modification of one, two and three Wilson coefficients at a time. To find an optimized constraint on $C_\phi$ from the data we perform a five-channel classification (two signal and three background modes including the $\hhbox$ contribution that is insensitive to modifications of $C_\phi$). To constrain either $C_{u\phi}$ or $C_{d\phi}$ we perform a four-channel classification taking the $gg$F channel as a single background mode. To constrain $C_\phi$ and one of $C_{u\phi}$ or $C_{d\phi}$ we perform a six-channel classification. Lastly, a simultaneous constraint on all three Wilson coefficients is realized with a seven-channel classification. All the codes and data necessary to reproduce the results we got from this interpretable machine learning framework are made available at a \texttt{Github} repository: \href{https://github.com/talismanbrandi/IML-diHiggs.git}{https://github.com/talismanbrandi/IML-diHiggs.git}.

To set the stage, we will define our measure of significance and how we estimate it. We first construct a confusion matrix from the predictions of the trained BDT. This is a $n\times n$ matrix, for $n$ channels. The sum of the elements in the $i^{th}$ row, $\sum_j N_{ij}$, gives the number of events produced in channel $i$ that would be generated in a pseudo-experiment with the projected luminosity corresponding to the actual experiment. The sum of the $j^{th}$ column, $\sum_i N_{ij}$, gives the number of events from channel $j$ predicted (including correct classifications and misclassifications) by the BDT in this pseudo-experiment. Hence, the $(i,j)$ element of the matrix gives the number of events of the $i^{th}$ class that is classified as belonging to the $j^{th}$ class with $i\ne j$ signifying a misclassification. The significance of the $j^{th}$ channel given by $S/\sqrt{S + B}$, $S$ being signal and $B$ being background, can be defined as
\begin{equation}
    \mathcal{Z}_j=\frac{|N_{jj}|}{\sqrt{\sum_i N_{ij}}},
\end{equation}
where $i$ is the row index and $j$ is the column index. For ease of interpretation, we will present our results also in terms of $\kappa_\lambda$, $\kappa_u$ and $\kappa_d$ which are simply the rescaling of the SM trilinear coupling and the light-quark Yukawa couplings of the up and down quarks, respectively.

\subsection{Constraints on light-quark Yukawa couplings at the HL-LHC and FCC-hh}
\label{sec:CqH}

The fact that machine learning algorithms can far outperform cut-and-count analyses is a bygone conclusion. Preliminary estimates of the HL-LHC reach for SM Higgs pair production can be found in Ref.~\cite{Cepeda:2019klc} and range from 4$\sigma$ to $4.5\sigma$ signal significance combining several channels and combining the ATLAS and CMS measurements. The $\bbaa$ final state alone allows for a $\sim2.7\sigma$ measurement. In Ref.~\cite{Alves:2017ued}, a more refined machine learning procedure using Bayesian Optimization has been suggested and it has been shown that, indeed, the measurement of a di-Higgs signal can be further improved over preliminary estimates made by ATLAS and CMS using the $\bbaa$ final state alone. A sensitivity of about $5\sigma$ can be achieved using their procedure with the caveat that they use $S/\sqrt{B}$ as the definition of significance with a very low number of correctly classified signal and background events. As an exercise, we repeated the BDT analysis with our framework and estimated a $\sim3.4\sigma$ signal significance for SM Higgs pair production, which is similar to the estimate made in Ref.~\cite{Alves:2017ued} without using any optimization. 

A better portrayal of the advantages gained by using a multivariate analysis can be made by comparing the constraints set on $C_{u\phi}$, or $\kappa_u$, and $C_{d\phi}$, or $\kappa_d$, from a cut-and-count (CC) analysis and a multivariate (MV) analysis allowing for the variation of only one Wilson coefficient at a time. The projected $1\sigma$ bounds at HL-LHC for 6$\inab$ of luminosity for a CC analysis are given in  Ref.~\cite{Alasfar:2019pmn} and compared to our results as follows
\begin{eqnarray}
    C_{u\phi}^{MV} \left(\kappa_u^{MV}\right) = [-0.09, 0.10] \;([-466, 454]),\quad C_{u\phi}^{CC} (\kappa_u^{CC}) = [-0.18, 0.17] \;([-841, 820]), \nonumber\\
    C_{d\phi}^{MV} (\kappa_d^{MV}) = [-0.16, 0.16] \;([-360, 360]),\quad C_{d\phi}^{CC} (\kappa_d^{CC}) = [-0.18, 0.18] \;([-405, 405]). \nonumber\\
\end{eqnarray}
From this, we clearly see a factor of $\sim$2 improvement in the bounds on $C_{u\phi}$ and $\mathcal O(10\%)$ improvement in the determination of $C_{d\phi}$. The projected bounds on these operators at FCC-hh with 30$\inab$ of data using our framework are
\begin{equation}
\begin{split}
    C_{u\phi}^{MV} \left(\kappa_u^{MV}\right) = [-0.012, 0.011] \;([-57.8, 54.7])\,,\\
    C_{d\phi}^{MV} (\kappa_d^{MV}) = [-0.012, 0.012] \;([-26.3, 28.4])\,.
\end{split}
\end{equation}
These projected bounds for FCC-hh are an order of magnitude better than those for HL-LHC. In addition, the bounds on $C_{u\phi}$ and $C_{d\phi}$ are numerically the same displaying a much greater improvement in the bounds on $C_{d\phi}$ than on $C_{u\phi}$ at the higher energy collider.

\subsection{Constraints on Higgs trilinear self-coupling at the HL-LHC and FCC-hh}
\label{sec:CH}
	
\begin{table}[]
    \centering
    {\footnotesize
    \begin{tabular}{ll|rrrrr|r}
    \multirow{7}{*}{\rb{\bf Actual no. of events\hspace{0.45cm}}} & \multicolumn{7}{c}{\bf Predicted no. of events at HL-LHC}\\
    \cmidrule[\heavyrulewidth]{2-8}
    & Channel & $\hhtri$ & $\hhtri$ &  $\hhbox$&      $\QQh$ & $\bbaa$ &   total \\
    \cline{2-8}
    &$\hhtri$         &   28&	14&	    18&	    38&	10&	        108     \\
    &$\hhint$         &   89&	80&	    129&	178&	41&	        517     \\
    &$\hhbox$         &   77&	105&	266&	265&	50&	        763     \\
    &$\QQh$           &  177&	98&	    191&	5,457&	1,835&      7,758   \\
    &$\bbaa$          &1,743&	845&	1,074&  30,849&	287,280&	321,791 \\
    \cline{2-8}
    &$\mathcal{Z}_j$& 0.61&	    2.37&	6.49&	28.45&	534.1&              \\
    \cmidrule[\heavyrulewidth]{2-8}
    \end{tabular}
    }
    \caption{\it Trained BDT classification (confusion matrix) of the five channels used to extract constraints on $C_\phi$ (or $\kappa_\lambda$) at HL-LHC with 6 $\iab$ luminosity (ATLAS+CMS), assuming SM signal injection. The right-most column gives the total number of events expected in each channel in the SM and the bottom-most row gives the signal significance.}
    \label{tab:HL-LHC-confusion-CH}
    
    \vspace{0.75cm}
    {\footnotesize
    \begin{tabular}{ll|rrrrr|r}
    \multirow{7}{*}{\rb{\bf Actual no. of events\hspace{0.45cm}}} & \multicolumn{7}{c}{\bf Predicted no. of events at FCC-hh}\\
    \cmidrule[\heavyrulewidth]{2-8}
    & Channel & $\hhtri$ & $\hhtri$ &  $\hhbox$&      $\QQh$ & $\bbaa$ &   total \\
    \cline{2-8}
    &$\hhtri$       &  	3,579&  1,303&	2,372&	4,697&	    337&	    12,288      \\
    &$\hhint$       &  13,602&  7,300&	17,075&	24,716&	    1523&	    64,216      \\
    &$\hhbox$       &  14,534&	11,416&	35,988&	415,26&     1,996&	    105,460     \\
    &$\QQh$         &  29,611&	12,355&	23,279&	1,238,266&	214,564&	1,518,075   \\
    &$\bbaa$        &  45,574&	22,290&	26,213&	150,935&	227,142&	24,317,657  \\
    \cline{2-8}
    &$\mathcal{Z}_j$&   10.95&	31.22&	111.1&	737.7&	    4,743&	 \\
    \cmidrule[\heavyrulewidth]{2-8}
    \end{tabular}
    }
    \caption{\it Trained BDT classification (confusion matrix) of the five channels used to extract constraints on $C_\phi$ (or $\kappa_\lambda$) at FCC-hh with 30 $\iab$ luminosity, assuming SM signal injection. The right-most column gives the total number of events expected in each channel in the SM and the bottom-most row gives the signal significance.}
    \label{tab:FCC-hh-confusion-CH}
\end{table}
\begin{figure}[t!]
	\centering
	\includegraphics[width=0.47\linewidth]{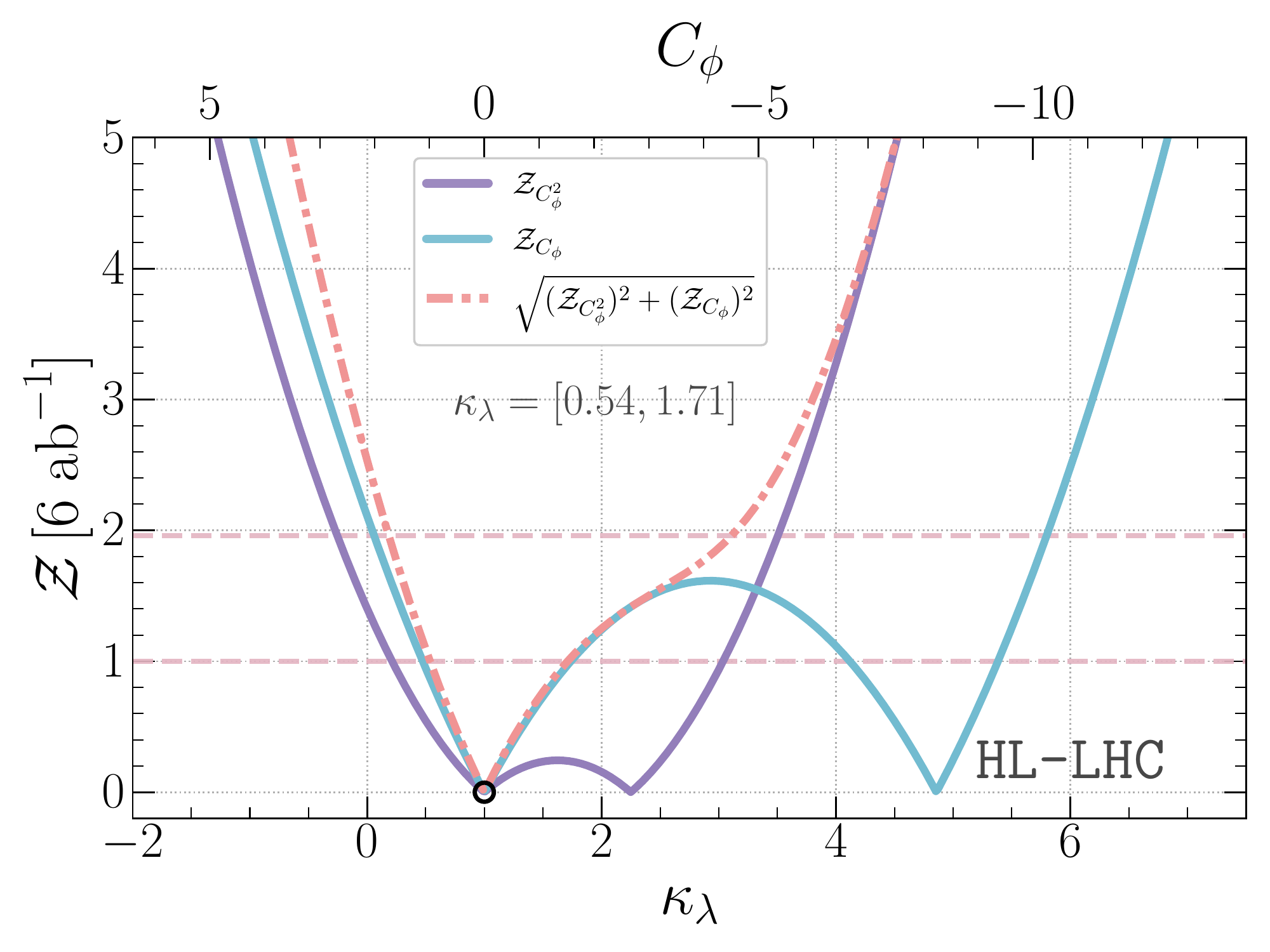}
	\includegraphics[width=0.47\linewidth]{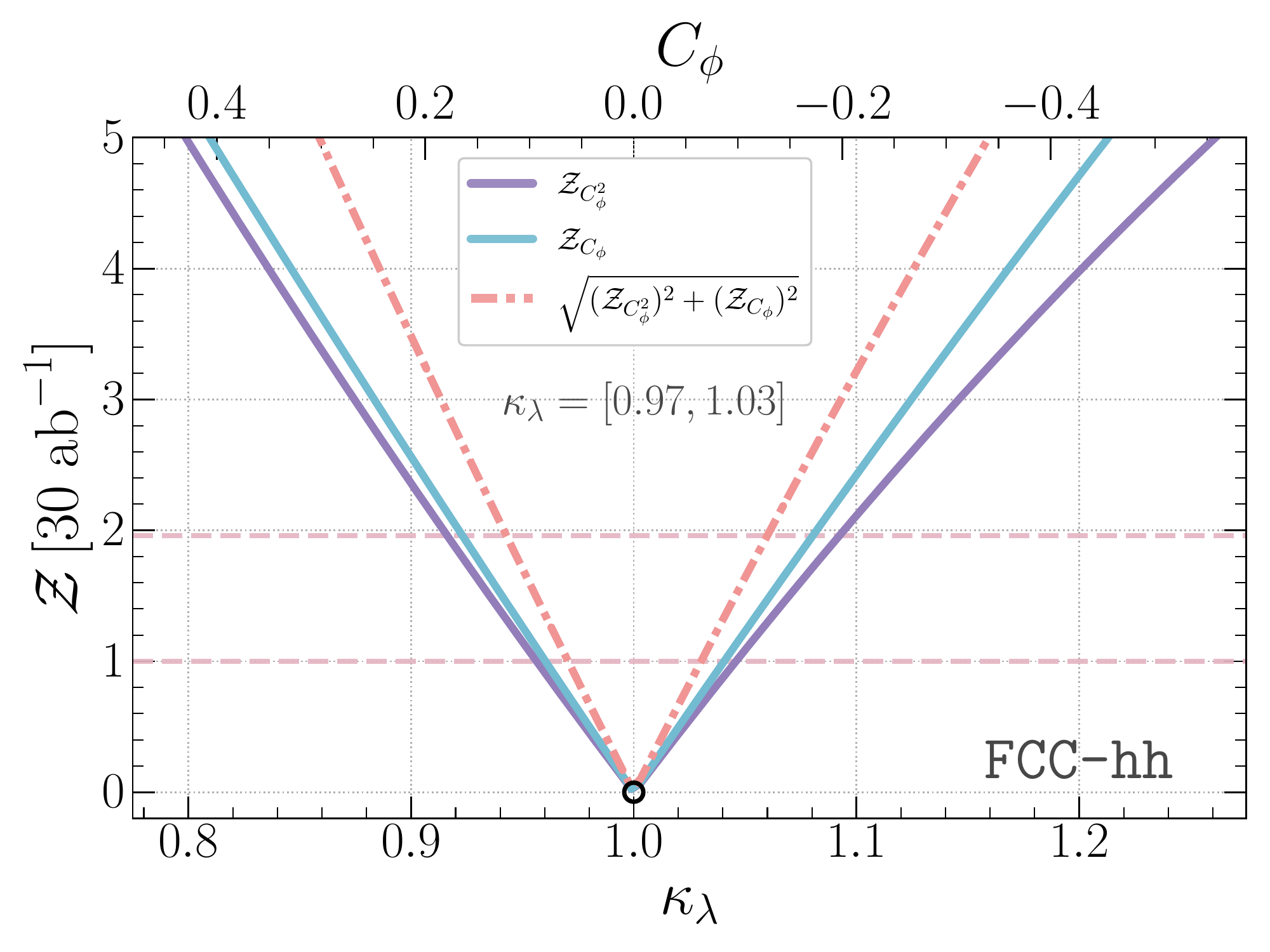}
	\caption{\it Bounds on $\kappa_\lambda$ (or $C_\phi$) at the HL-LHC (left panel) and the FCC-hh (right panel). The solid blue lines are the constraints coming from the $\hhint$ contribution which scales linearly with the modified coupling and the solid purple line is that from the $\hhtri$ contribution that scales quadratically with the modified coupling. The red dot-dashed line is the combination of the quadratic and linear channel. The horizontal light red dashed lines mark the 68\% and 95\% confidence intervals. The 68\% CL bounds on $\kappa_\lambda$ are given within square bracket.}
	\label{fig:constraintkl}
\end{figure}

In \autoref{tab:HL-LHC-confusion-CH}, we provide the output of the BDT classification for 6 $\iab$ of data collected at HL-LHC and in \autoref{tab:FCC-hh-confusion-CH}, we provide the same for 30 $\iab$ of data at FCC-hh. It can be seen from these matrices that while the $\bbaa$ QCD-QED channel is the dominant background, the BDT performs better in separating it from the signal channels than separating $\QQh$. This is due to the kinematic similarities between the signal and the $\QQh$ background.

In \autoref{fig:constraintkl}, we present the constraints on $\kappa_\lambda$ (or $C_\phi$) that can be set from HL-LHC in the left panel and FCC-hh in the right panel. The $\hhbox$ topology is not modified by $C_\phi$ and serves as a background to the measurement of $C_\phi$. We separate the constraints from the $\hhtri$, which is quadratic in $C_\phi$ from the $\hhint$ which is linear in $C_\phi$. The combination of the two is given by the red dot-dashed line and is asymmetric around the best-fit point, for SM signal injection, $\kappa_\lambda=1$\; $(C_\phi=0)$. The projected $1\sigma$ bound on $\kappa_\lambda$ is $[0.54, 1.71]$ at HL-LHC. There is a vast improvement projected for the FCC-hh which is not only due to increased luminosity but also due to the measurement being at a higher energy. The projected $1\sigma$ bound on $\kappa_\lambda$ is about 3\%.

\subsection{Two and three parameter constraints on Higgs couplings}
\label{sec:multiparam}

{The primary focus of this work is to move beyond just looking at constraints on $C_\phi$ from Higgs pair production and to shed light on how simultaneous modifications of the light-quark Yukawa couplings due to non-zero contributions from $C_{u\phi}$ and $C_{d\phi}$ can change the \unskip\parfillskip 0pt \par}

\begin{figure}[t!]
	\centering
	\includegraphics[width=0.47\linewidth]{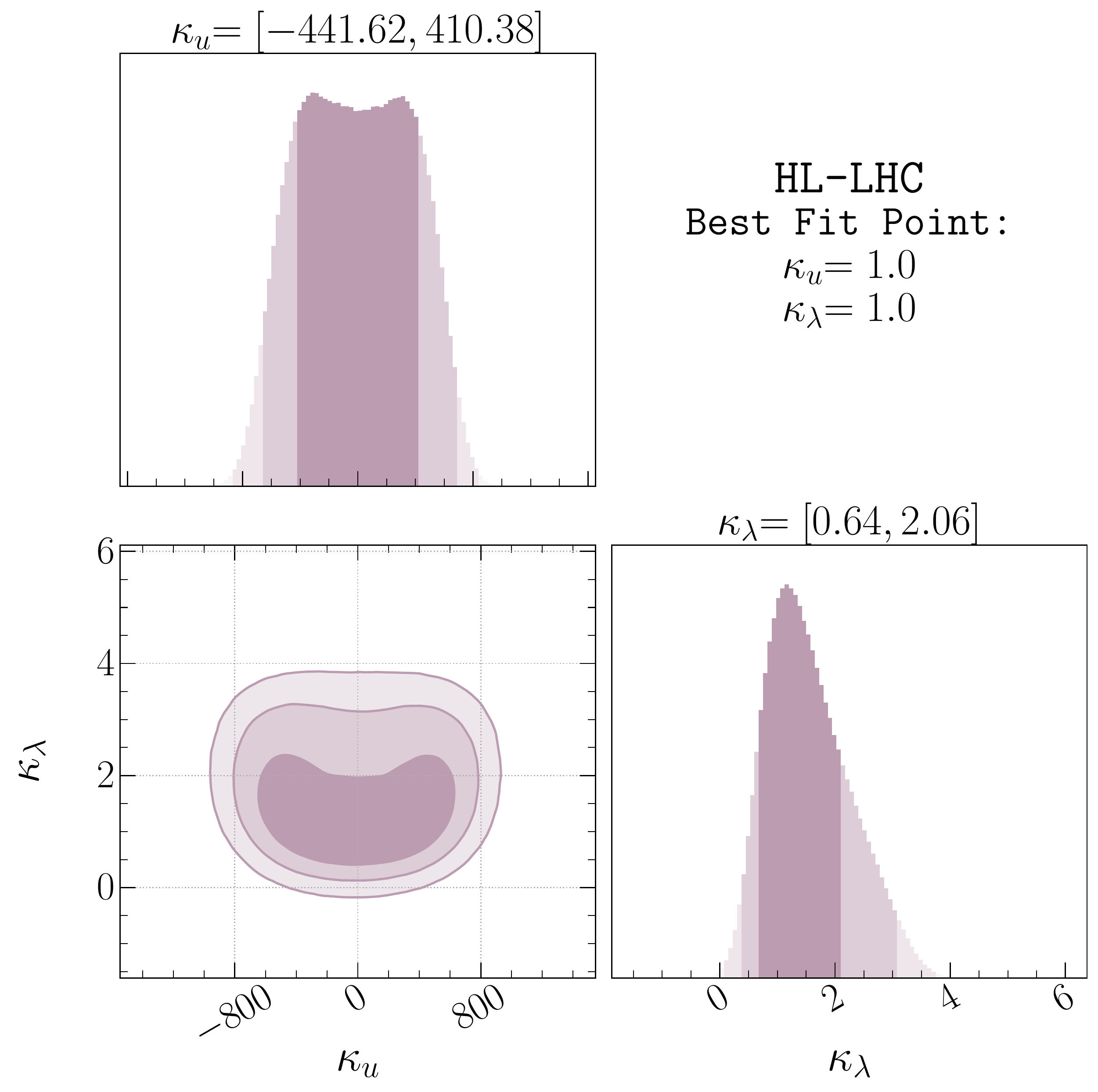}
	\includegraphics[width=0.47\linewidth]{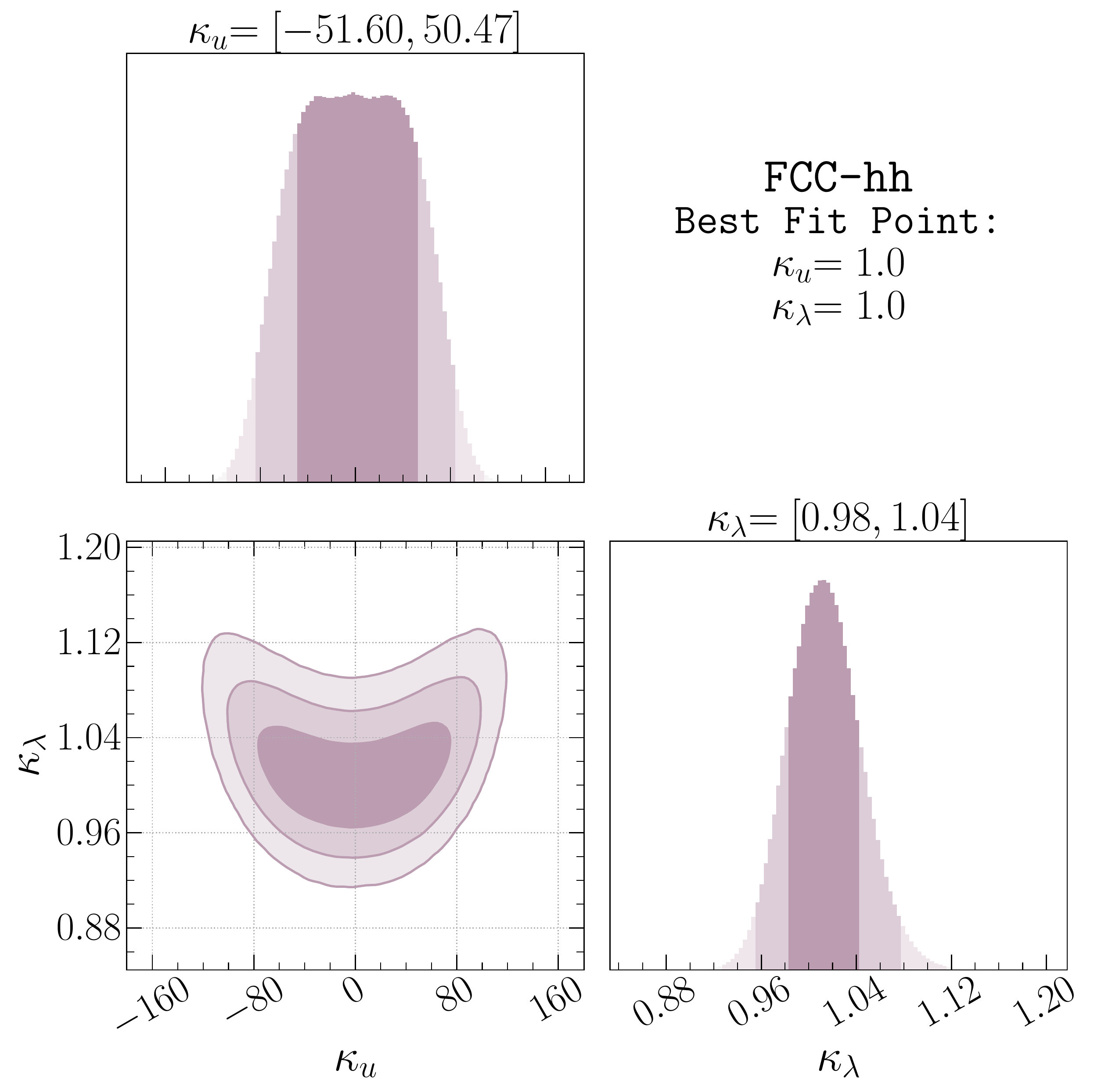}
	\includegraphics[width=0.47\linewidth]{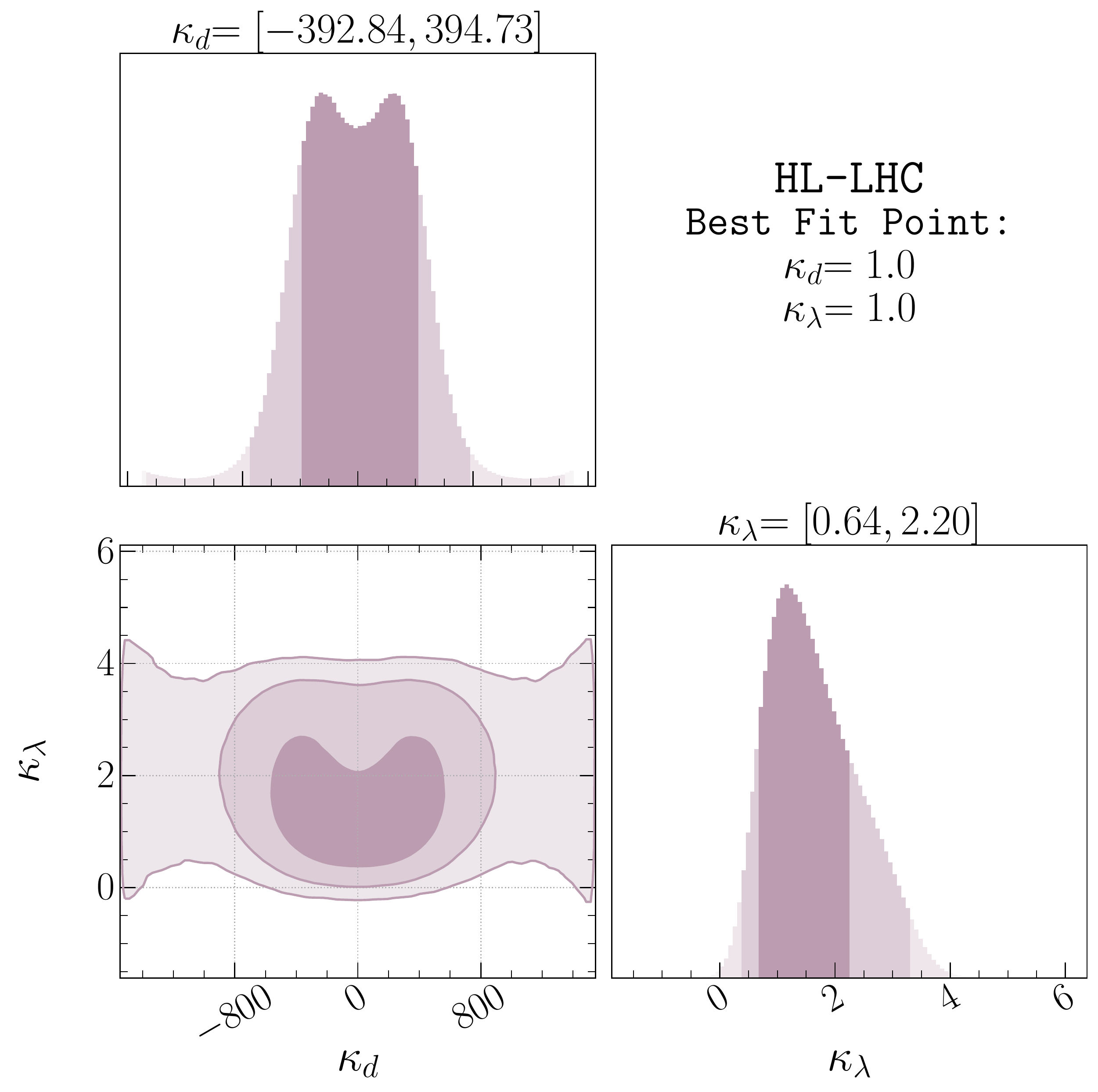}
	\includegraphics[width=0.47\linewidth]{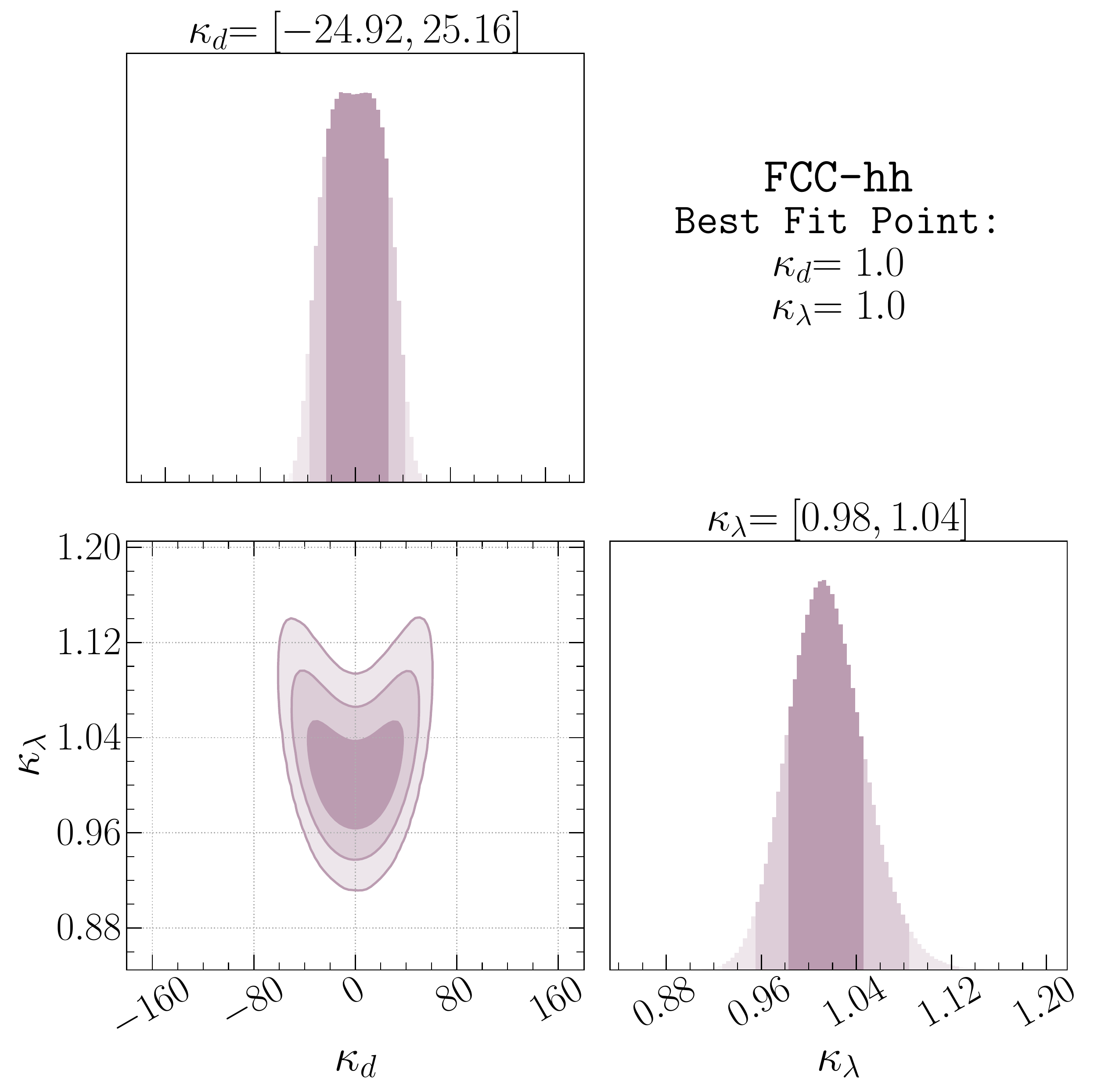}
	\includegraphics[width=0.47\linewidth]{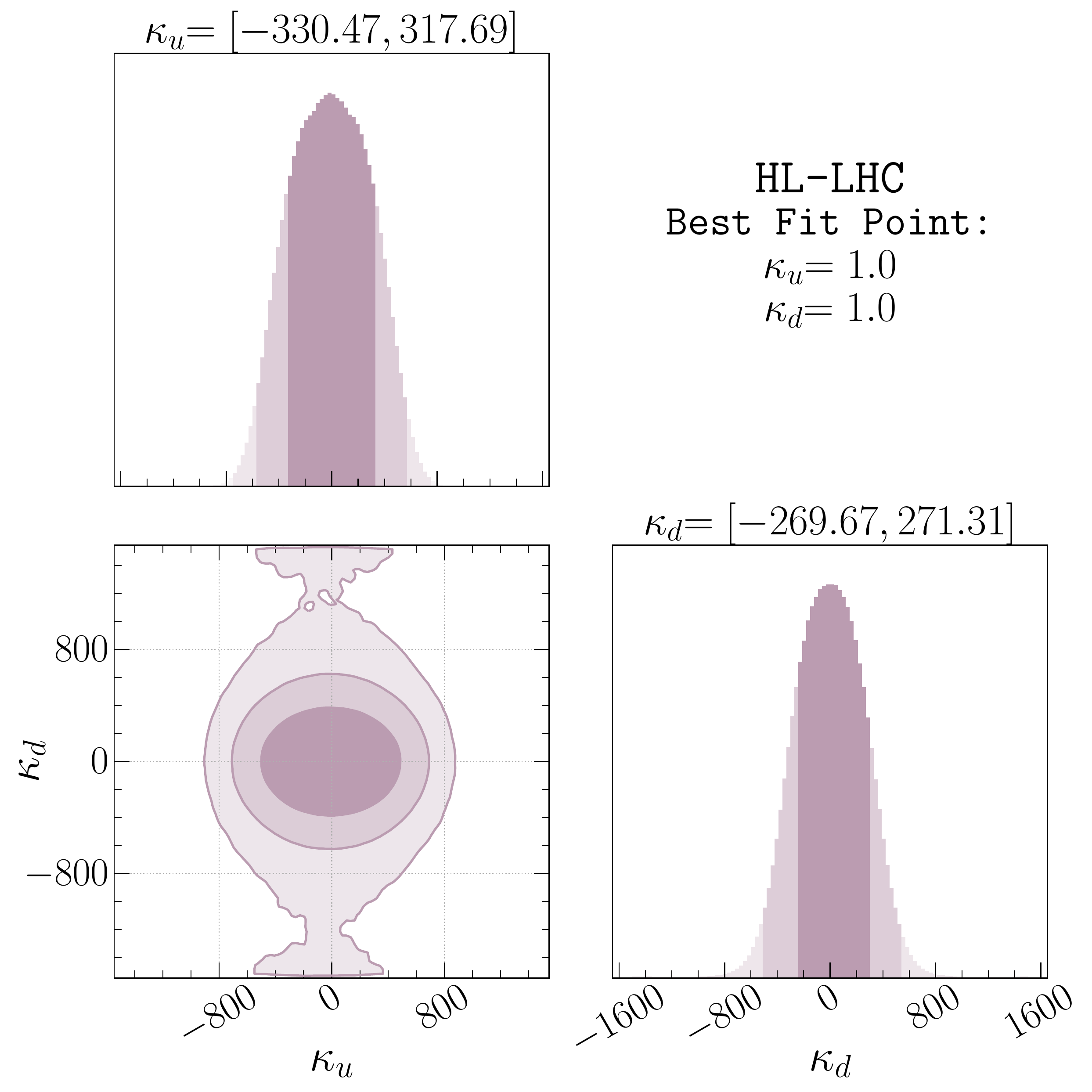}
	\includegraphics[width=0.47\linewidth]{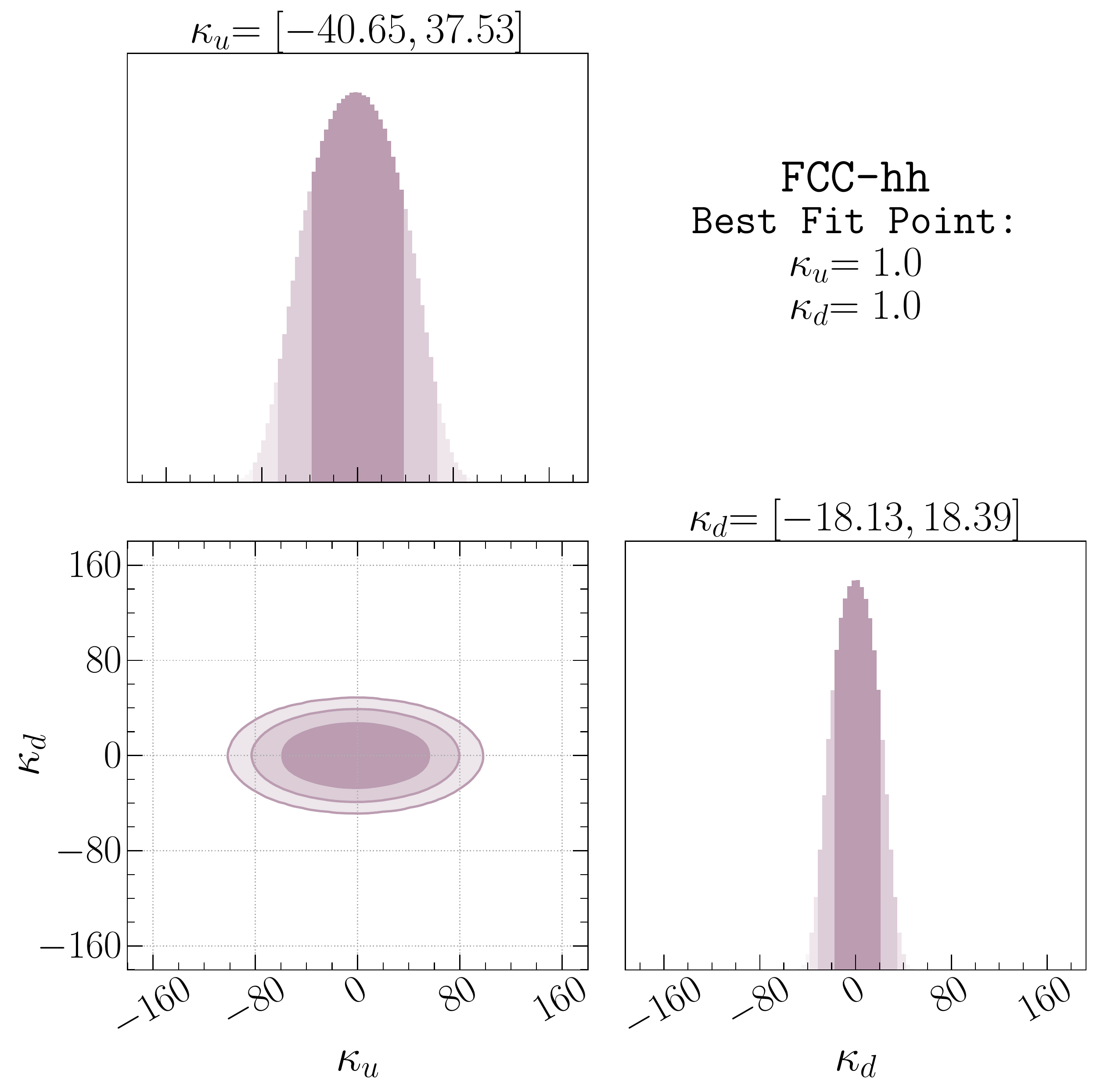}
	\caption{\it Constraints on pairs of Wilson coefficients for $\kappa_\lambda$, $\kappa_{u}$ and $\kappa_{d}$, The panels on the left are for HL-LHC with 6 $\iab$ of luminosity and the ones on the right are for FCC-hh with 30 $\iab$ of luminosity. The one-dimensional distributions are derived from the two-dimensional distributions by marginalization using uniformly distributed priors for the independent variables with ranges sufficiently large to accommodate much more than 5$\sigma$ variation of the variables from their central values.}
	\label{fig:constraint2d}
\end{figure}
\FloatBarrier

\noindent constraints on $C_\phi$. The modifications of the light-quark Yukawa couplings manifest themselves in two different ways. Firstly, non-zero $C_{u\phi}$ and $C_{d\phi}$ open up the $q\bar q \to hh$ production mode through a contact interaction (see \autoref{fig:qqa}) thus changing the production cross-section of the di-Higgs channel. This increase in the production cross-section sets the tightest constraints on $C_{u\phi}$ and $C_{d\phi}$ from Higgs pair production. Secondly, the modifications of the light-quark Yukawa couplings also modify the branching fraction of $h\to\gamma\gamma$ and the width of the Higgs boson. The latter modifies the channels that are also sensitive to $C_\phi$, thus modifying the constraints that can be set on $C_\phi$ from future measurements. Such constraints from these modifications are the subdominant ones on $C_{u\phi}$ and $C_{d\phi}$ but they need to be taken into account for a holistic picture.  

\begin{table}[t!]
    \centering
    {\footnotesize
    \begin{tabular}{cccc||ccc}
    \toprule
         Operators &  $C_{u\phi}$&   $C_{d\phi}$&   $C_{\phi}$&    $\kappa_{u}$&   $\kappa_{d}$&   $\kappa_\lambda$\\
    \midrule
         \multicolumn{7}{c}{HL-LHC 14 TeV 6$\inab$}\\
    \midrule
         $\mathcal O_{\phi}$ &--   & --            &[-1.6, 1.0] &--  & -- &[0.53, 1.7]\\
         $\mathcal O_{u\phi}$&[-0.09, 0.10]   & --            &-- &[-480, 430]  & -- &--\\
         $\mathcal O_{d\phi}$&--   & [-0.16, 0.16]            &-- &--  & [-360, 360] &--\\
         $\mathcal O_{u\phi}$ \& $\mathcal O_{\phi}$ &[-0.087, 0.091]  & --            &[-2.4, 0.79] &[-430, 420] & -- &[0.63, 2.1]\\
         $\mathcal O_{d\phi}$ \& $\mathcal O_{\phi}$ & --             &[-0.17, 0.17]  &[-2.7, 0.77]& -- &[-380, 380] &[0.63, 2.3]\\
         $\mathcal O_{u\phi}$ \& $\mathcal O_{d\phi}$&[-0.066, 0.069]&[-0.12, 0.12]& --            &[-330, 310] &[-270, 270] & -- \\
         All                                   &[-0.077, 0.084]&[-0.16, 0.16]& [-2.8, 0.43]&[-400, 370] &[-360, 360] & [0.79, 2.3] \\
    \midrule
    \midrule
         \multicolumn{7}{c}{FCC-hh 100 TeV 30$\inab$}\\
    \midrule
         $\mathcal O_{\phi}$ &--   & --            &[-0.066, 0.064] &--  & -- &[0.97, 1.03]\\
         $\mathcal O_{u\phi}$&[-0.012, 0.011]   & --            &-- &[-58, 55]  & -- &--\\
         $\mathcal O_{d\phi}$&--   & [-0.012, 0.011]            &-- &--  & [-26, 28] &--\\
         $\mathcal O_{u\phi}$ \& $\mathcal O_{\phi}$ &[-0.010, 0.011]  & --            &[-0.091, 0.042] &[-52, 49] & -- &[0.98, 1.04]\\
         $\mathcal O_{d\phi}$ \& $\mathcal O_{\phi}$ & --             &[-0.010, 0.012]  &[-0.092, 0.041]& -- &[-24, 26] &[0.98, 1.04]\\
         $\mathcal O_{u\phi}$ \& $\mathcal O_{d\phi}$&[-0.008, 0.009]&[-0.008, 0.009]& --            &[-42, 39] &[-19,19] & -- \\
         All                                   &[-0.009, 0.010]&[-0.009, 0.010]& [-0.11, 0.023]&[-47, 44] &[-21, 21] & [0.99, 1.05] \\
    \bottomrule
    \end{tabular}
    }
    \caption{\it The 1$\sigma$ bounds on $C_{u\phi}$, $C_{d\phi}$ and $C_\phi$ from one-, two- and three-parameter fits for HL-LHC with 6 $\inab$ of data and FCC-hh with 30 $\inab$ of data. The corresponding bounds on the rescaling of the effective couplings, $\kappa_u$, $\kappa_d$ and $\kappa_\lambda$ are presented on the right side of the table.}
    \label{tab:twoparambounds}
\end{table}

In the two-parameter fits, we consider three possible scenarios. Firstly, one can assume that the trilinear Higgs coupling is not modified and only the light-quark Yukawa couplings are. Two other possibilities are the simultaneous modification of the $C_\phi$ and one of $C_{u\phi}$ and $C_{d\phi}$. These are the three constraints that we show in \autoref{fig:constraint2d} in terms of $\kappa_\lambda$, $\kappa_u$  and $\kappa_d$ respectively. As before, the constraints have been obtained by training the BDT to separate the relevant signal channels from the background, the signal used being the one corresponding to the set of Wilson coefficients that we wish to constrain. The confusion matrices for all three cases can be found in the \texttt{Github} repository (\href{https://github.com/talismanbrandi/IML-diHiggs.git}{https://github.com/talismanbrandi/IML-diHiggs.git}) for this analysis. The left panels of \autoref{fig:constraint2d} show the projected constraints for HL-LHC and the right panels for the FCC-hh.

Comparing with the constraints on $\kappa_\lambda$ given in \autoref{sec:CH} and \autoref{fig:constraintkl}, it can be seen from the top and middle left panels of \autoref{fig:constraint2d} that, indeed, the constraints on $\kappa_\lambda$ are diluted when the light-quark Yukawa couplings are allowed to vary. This effect is somewhat more prominent for $\kappa_{d}$ than for $\kappa_{u}$. This distinction stems from the fact that away from $\kappa_{u,d} = 1$ larger negative values of $\kappa_\lambda$ are allowed by the crescent-shaped curves in \autoref{fig:constraint2d}. For $\kappa_{d}$ vs.~$\kappa_\lambda$ the 3$\sigma$ region is unbounded in the domain $|\kappa_{d}| \gtrsim 1000$. The bounds on $\kappa_{u}$ and $\kappa_{d}$ from the fits with two parameters including $\kappa_\lambda$ remain the same as the bounds on these Wilson coefficients from the single parameter $\kappa_{u,d}$ fits. We summarize the results in~\autoref{tab:twoparambounds}.

It should be noted that the two-parameter fit for $\kappa_{u}$ and $\kappa_{d}$ provide a stronger bound on the two parameters than the fit performed individually. While this might be a bit counter-intuitive considering constraints from fits tend to deteriorate with the increasing number of parameters, we found that is not the case here. The reason is that the two-parameter fit is performed with the predictions made by the BDT trained with simulated events for both $\uuA$ and $\ddA$. Between these two channels, each forms the background for the other when separating them through a confusion matrix. Since the training also gives the proportion of mistagged events, both the signal and the backgrounds are modified by the Wilson coefficients leading to a greater deformation of the likelihood in a favourable direction such that the constraints on the Wilson coefficients in the two-parameter fit is better than for the case in which they were separated from other $\bbaa$ backgrounds individually.

\begin{figure}[t!]
	\centering
	\includegraphics[width =0.47\linewidth]{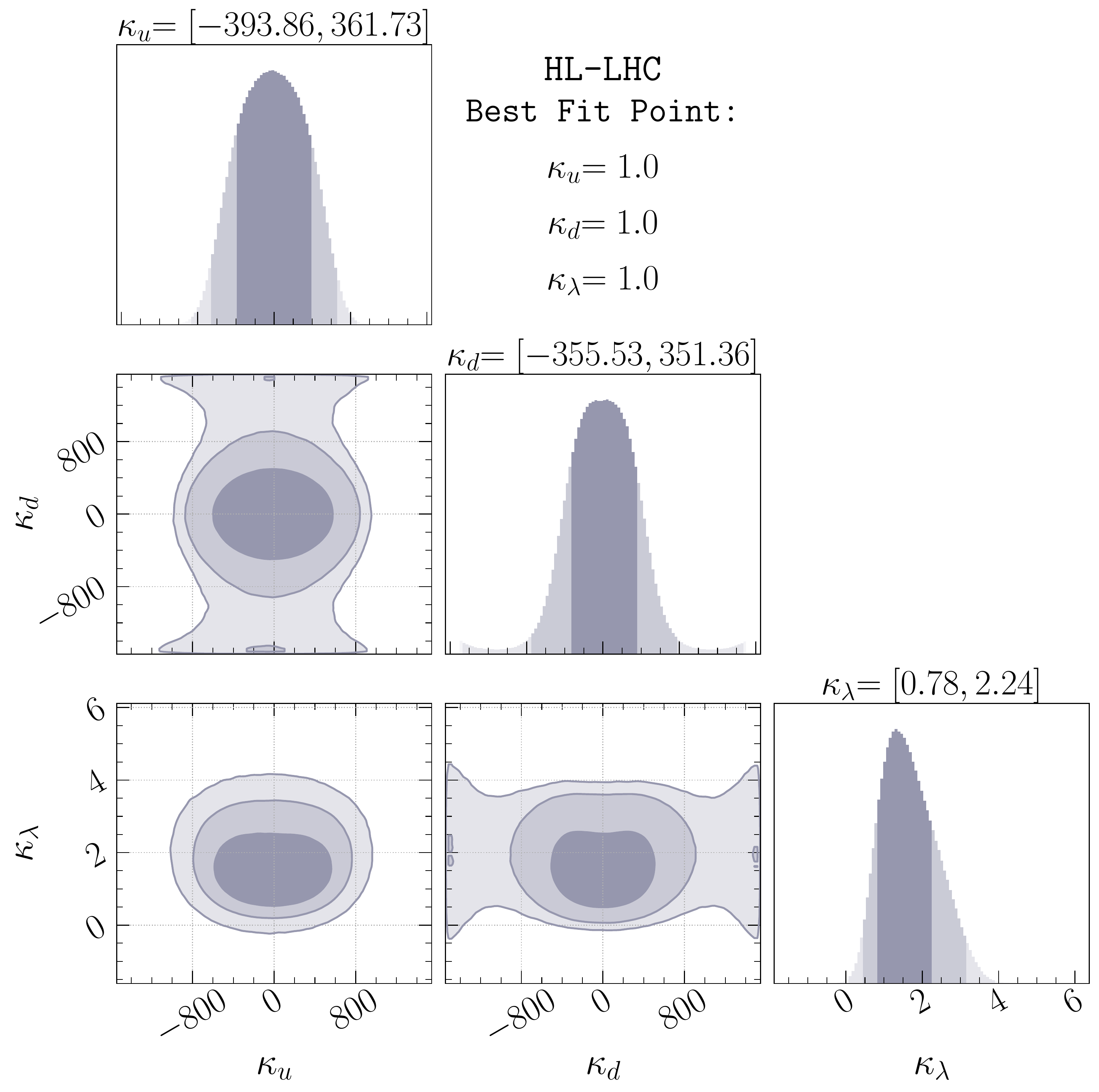}
	\includegraphics[width =0.47\linewidth]{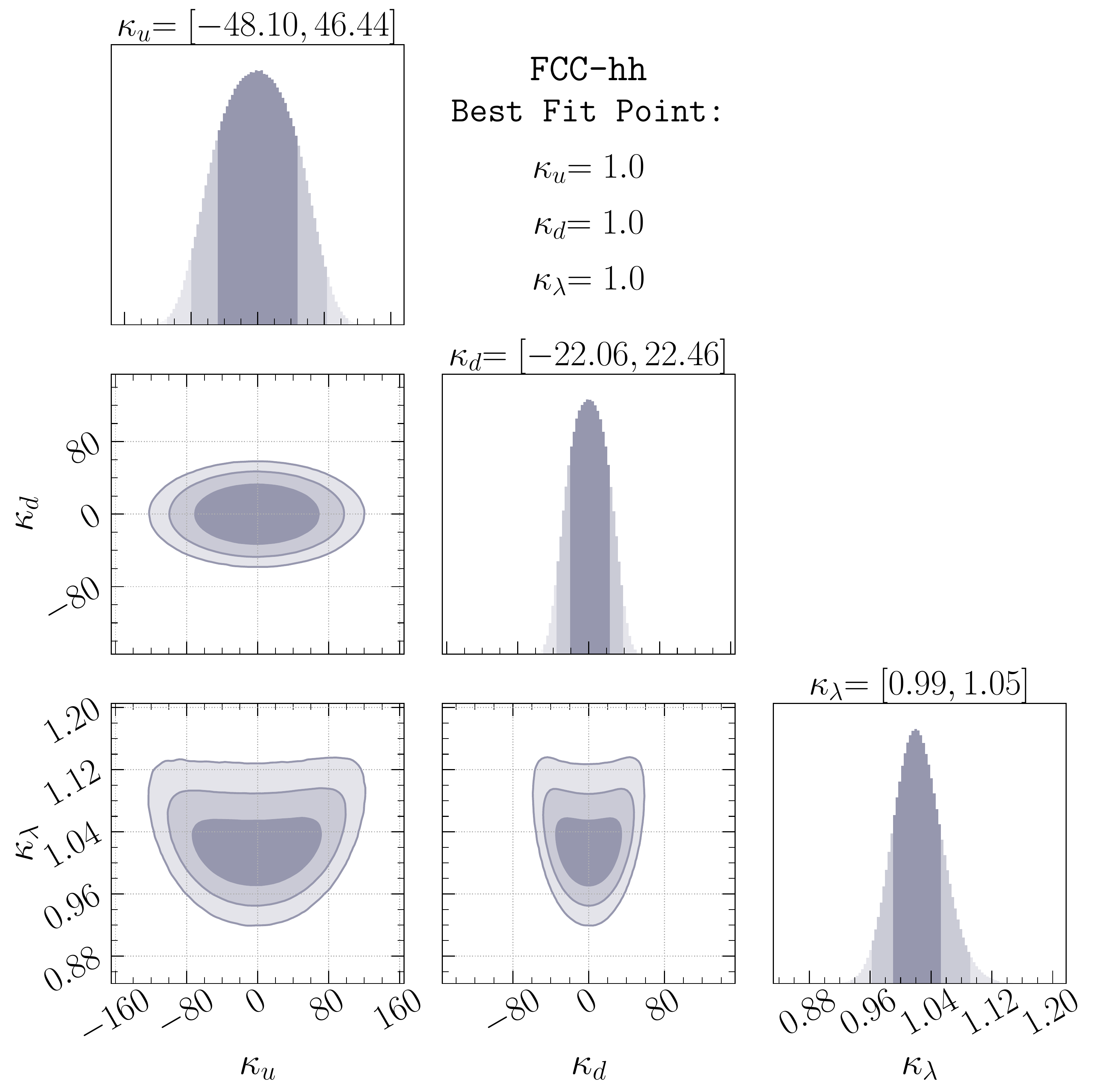}
	\caption{\it Three parameter fits with $C_{u\phi}$, $C_{d\phi}$ and $C_\phi$, 6 $\inab$ of luminosity at 14 TeV for HL-LHC (left panel) and 30 $\inab$ of luminosity at 100 TeV for FCC-hh (right panel). The one-dimensional distributions are derived from the two-dimensional distributions by marginalization using uniformly distributed priors for the independent variables with ranges sufficiently large to accommodate much more than 5$\sigma$ variation of the variables from their central values.}
	\label{fig:constraint3d}
\end{figure}

Finally, we perform a combined three-parameter fit including $\kappa_\lambda$, $\kappa_{u}$ and $\kappa_{d}$ ($C_\phi$, $C_{u\phi}$ and $C_{d\phi}$), with the results shown in~\autoref{fig:constraint3d}. For the same reason as explained before, the bounds on $\kappa_{u}$ and $\kappa_{d}$ are somewhat better than the two-parameter fits of these operators individually with $\kappa_\lambda$. The HL-LHC and FCC-hh projected bounds on $\kappa_\lambda$ remain nearly the same as those from the corresponding two-parameter fits. In \autoref{tab:twoparambounds} we also provide the bounds on $C_\phi$, $C_{u\phi}$ and $C_{d\phi}$ for comparison.

\subsection{Interpretation of Shapley values}
\label{sec:interp}

Finally, we want to demonstrate the interpretability of the machine learning framework we use and discuss the physics that allows for the separation of the signal channels from the background channels. The advantage of using an interpretable multivariate framework is that one can easily understand which of the kinematic variables are important for orchestrating this separation in a manner that significantly improves upon a cut-and-count analysis. We use a measure derived from Shapley values, $\overline{|S_v|}$, to understand the importance of each kinematic variable and understand the differences in kinematic shapes that separate the signal from the background.

\begin{figure}[t!]
	\centering
	\includegraphics[width=0.45\linewidth]{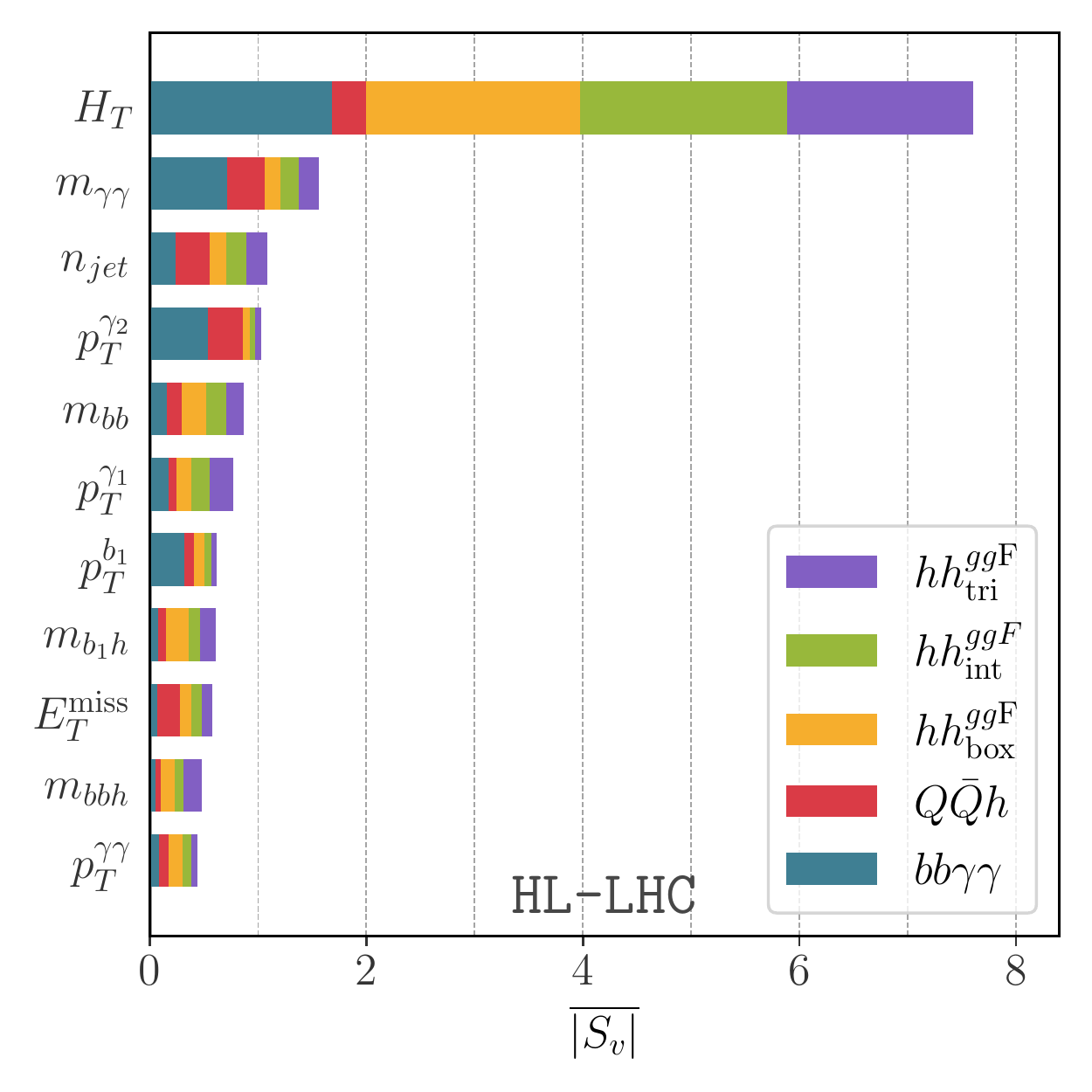}
	\includegraphics[width=0.45\linewidth]{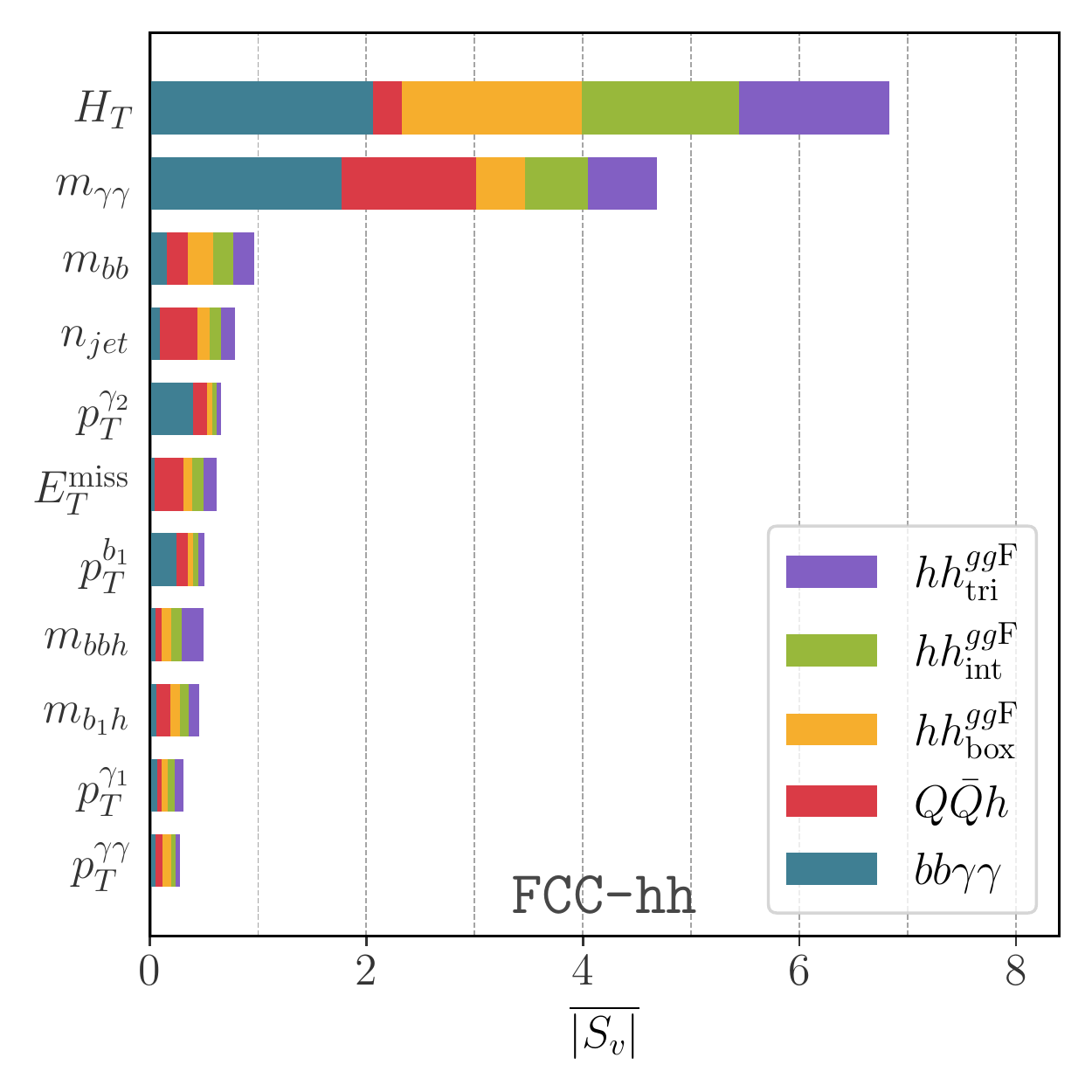}
	\includegraphics[width=0.45\linewidth]{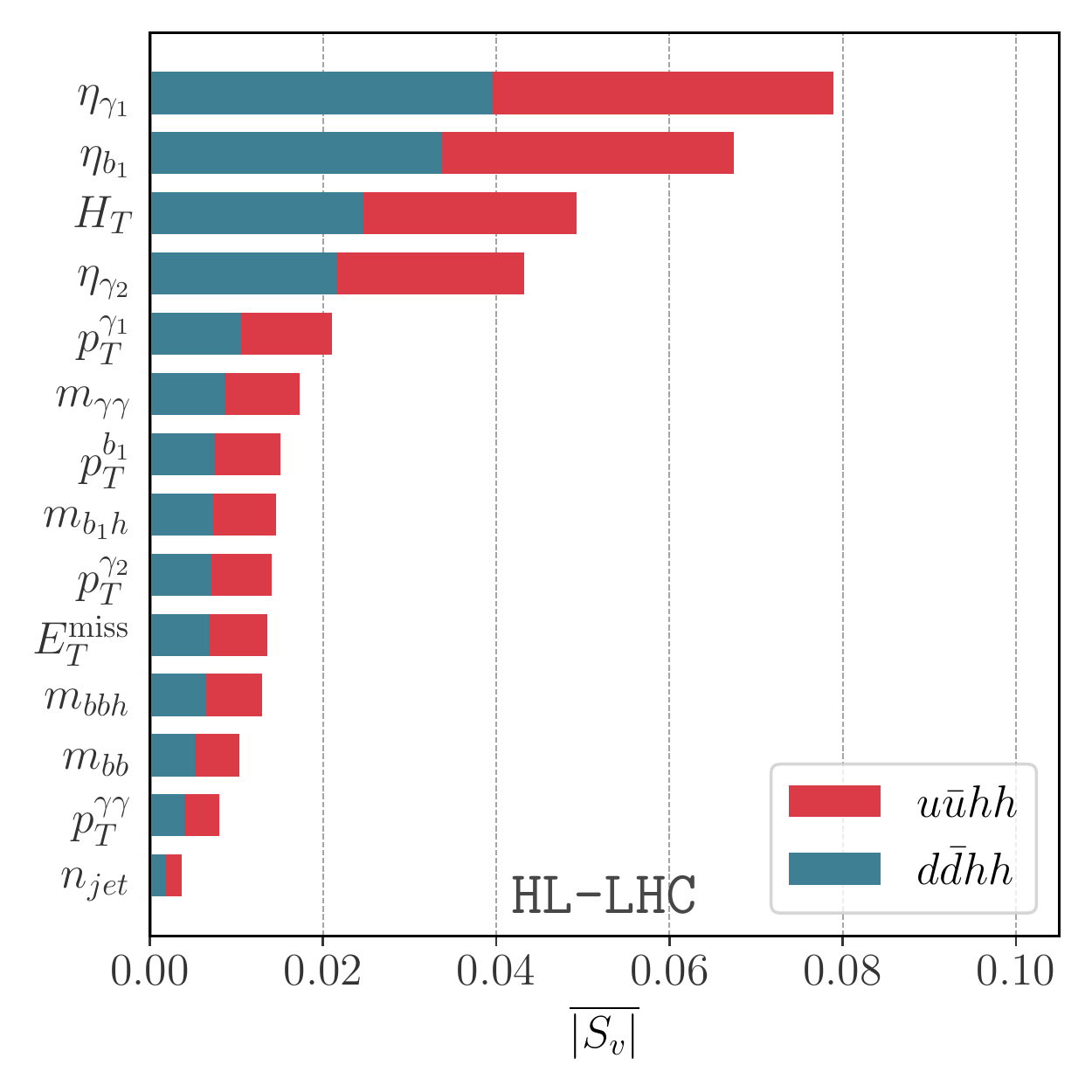}
	\includegraphics[width=0.45\linewidth]{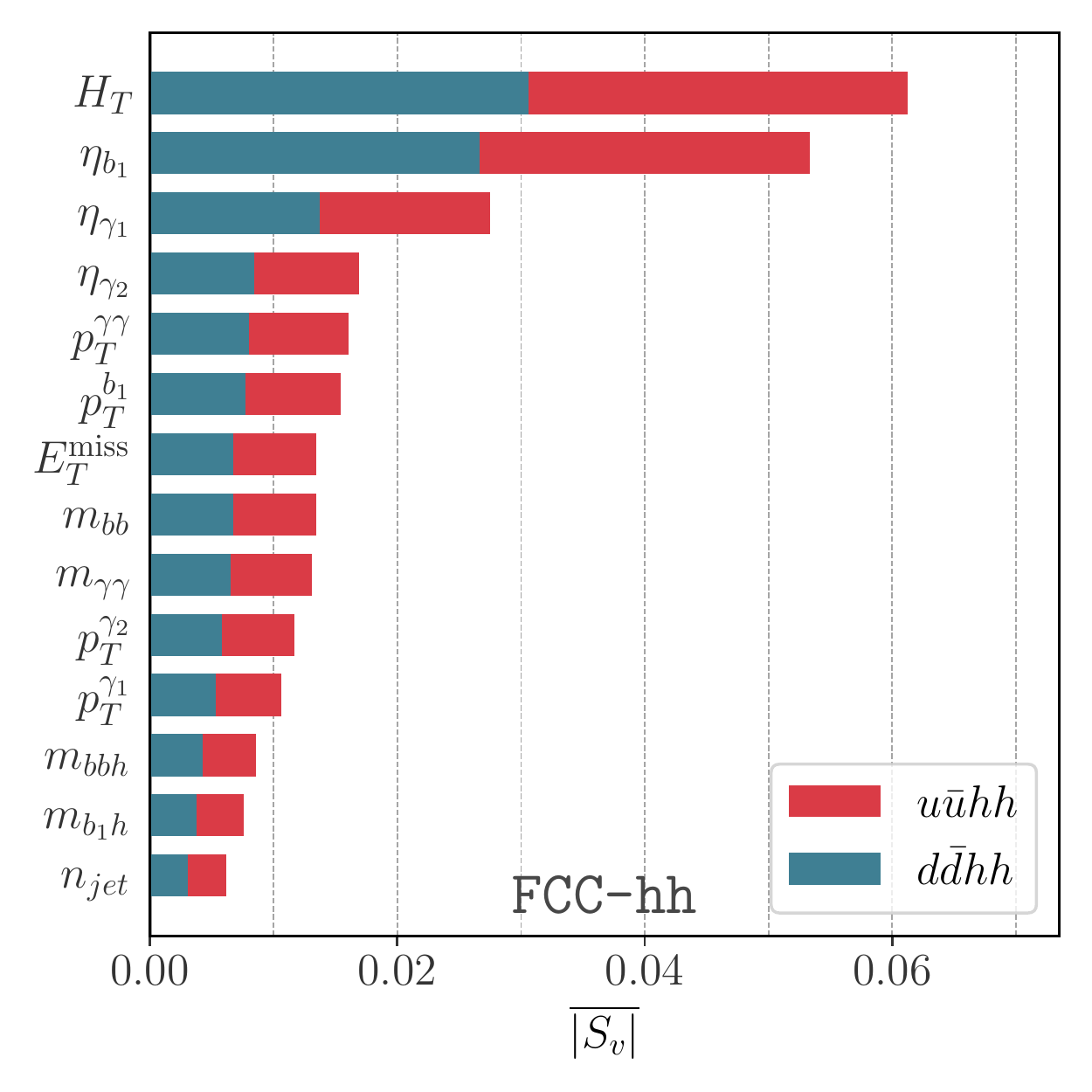}
	\includegraphics[width=0.45\linewidth]{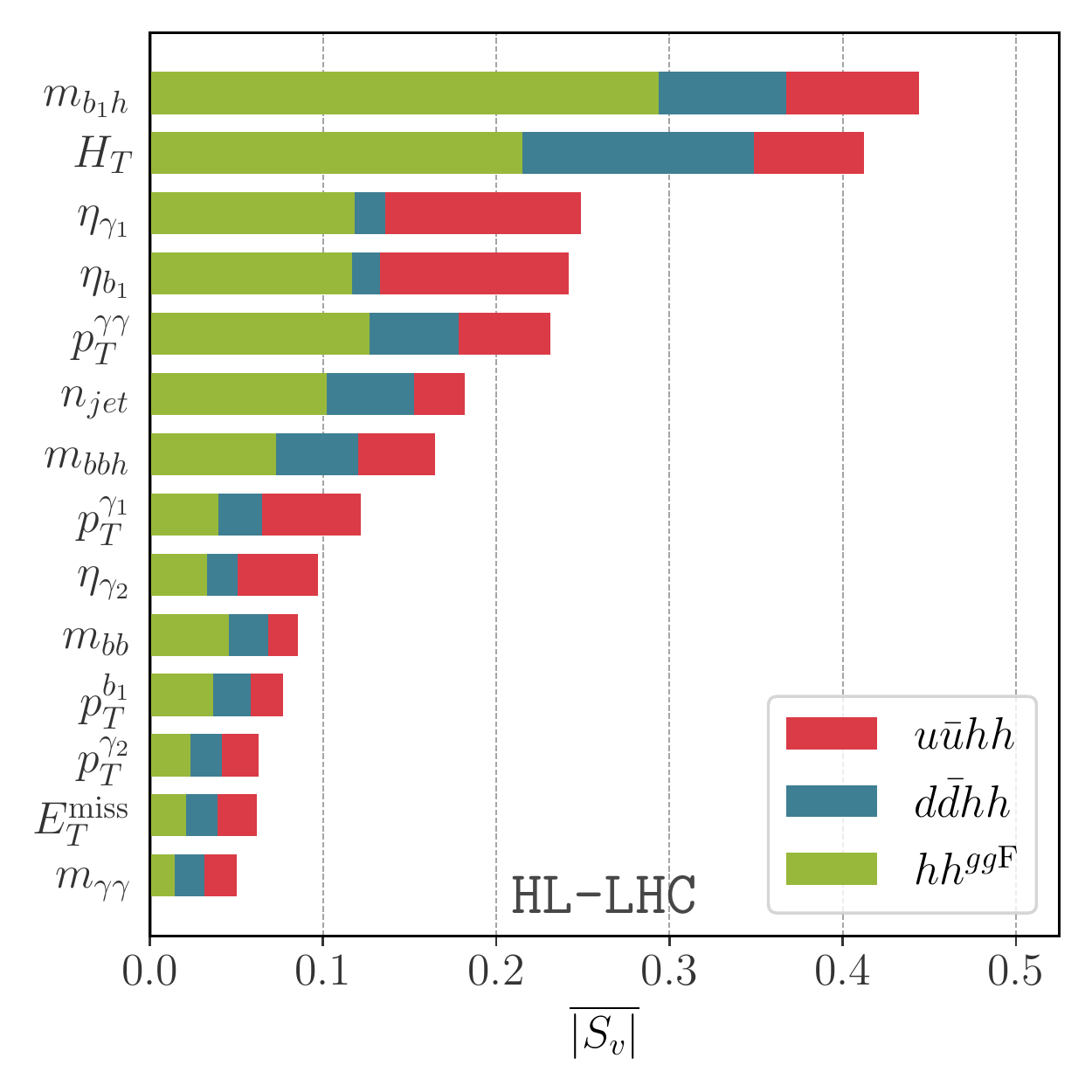}
	\includegraphics[width=0.45\linewidth]{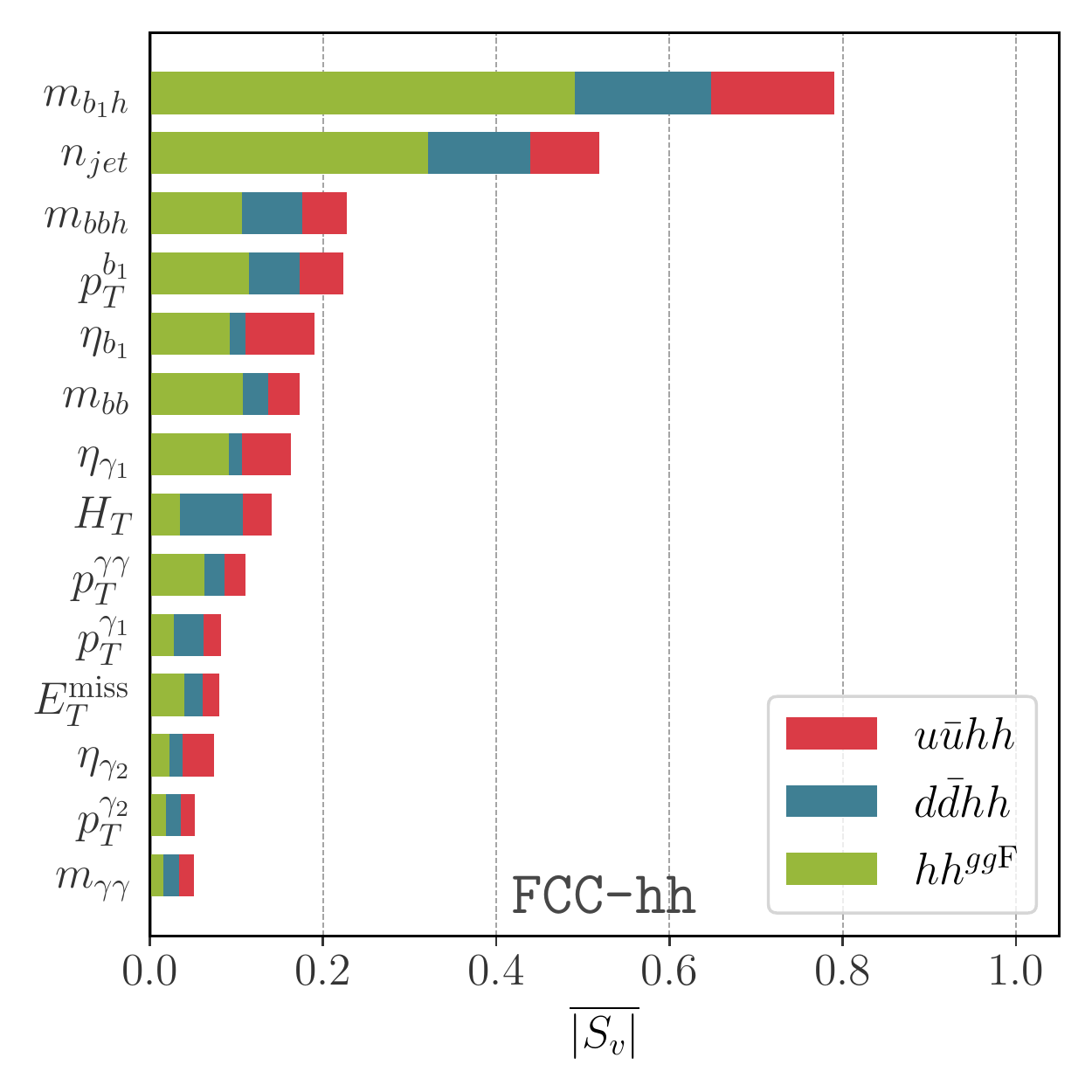}
	\caption{\it Top panels: The hierarchy of variables important for the separation of $\hhtri$ from $\hhint$ events from $\hhbox$, $\QQh$ and $\bbaa$ QCD-QED background at HL-LHC (left panel) and FCC-hh (right panel). Middle panels: The hierarchy of variables important for the separation of $\uuA$ from $\ddA$ events at HL-LHC (left panel) and FCC-hh (right panel). Lower panels: The hierarchy of variables important for the separation of $hh^{gg\rm F}$, $\uuA$ and $\ddA$ events at HL-LHC (left panel) and FCC-hh (right panel). The higher the value of $\overline{|S_v|}$ is, the more important the kinematic variable is in separating the different channels.}
	\label{fig:shap}
\end{figure}

{To give a feeling of what the values of $S_v$ mean, let us examine a single event. Assuming we have trained the BDT with $n$ kinematic variables and each event has $n\times m$ Shapley values associated with it, $m$ being the number of channels (signal and background channels). For a particular channel, $j$, and kinematic variable, $i$, $S_v$ can be positive or negative. A positive value implies that it is more likely that the event belongs to channel $j$ given the value of the kinematic variable $i$. Conversely, a negative value implies that the event is less likely to belong to channel $j$ given the value of the kinematic variable $i$. So regardless of whether $S_v$ is positive or negative, it helps in the sorting of events into various channels. Hence, $\overline{|S_v|}$ for a particular variable represents the strength of the variable to distinguish between channels. When summed over all channels this gives an overall picture of how good a discriminant a kinematic variable is for the processes involved. This is what is shown in \autoref{fig:shap} which we will now elaborate upon.

To begin with, we take a look at the $\overline{|S_v|}$ computed for the five-channel analysis performed for separating $\hhtri$ and $\hhint$ channels from $\hhbox$, $\QQh$ and $\bbaa$ QCD-QED background. In \autoref{fig:shap} we see the hierarchy plots for HL-LHC (top left panel) and FCC-hh (top right panel) generated from the predictions made by the BDT for this five-channel analysis. For both the colliders, $H_T$ is the most important variable that is bringing about the separation of the $\hhtri$ and $\hhint$ channels from the dominating $\bbaa$ QCD-QED background. The second most important variable is $m_{\gamma\gamma}$. The importance of $m_{\gamma\gamma}$ accentuates the separation of the background by a greater degree at FCC-hh than at HL-LHC.

For the separation between the two $\qqA$ channels, the story is very different. From the middle panels of \autoref{fig:shap} we see that the separation of $\uuA$ and $\ddA$ is truly a multivariate problem. Not surprisingly, the picture is very different for HL-LHC and FCC-hh. The differences between the two channels are driven by the differences in the parton distribution functions (PDF) of the up and down quarks. Since the PDF for the quarks change significantly from 14 TeV to 100 TeV, the variables that affect the separation of the two channels also change. Thus $\overline{|S_v|}$ give us a true picture of how distributions of several kinematic variables determine the separation of different channels that are mostly similar. When comparing the abscissa of the top two panels with the middle two panels one will also notice that $\overline{|S_v|}$ assumes much smaller values in the separation of $\uuA$ and $\ddA$. \unskip\parfillskip 0pt \par}

\FloatBarrier

\begin{figure}[h!]
	\centering
	\includegraphics[width=0.85\linewidth]{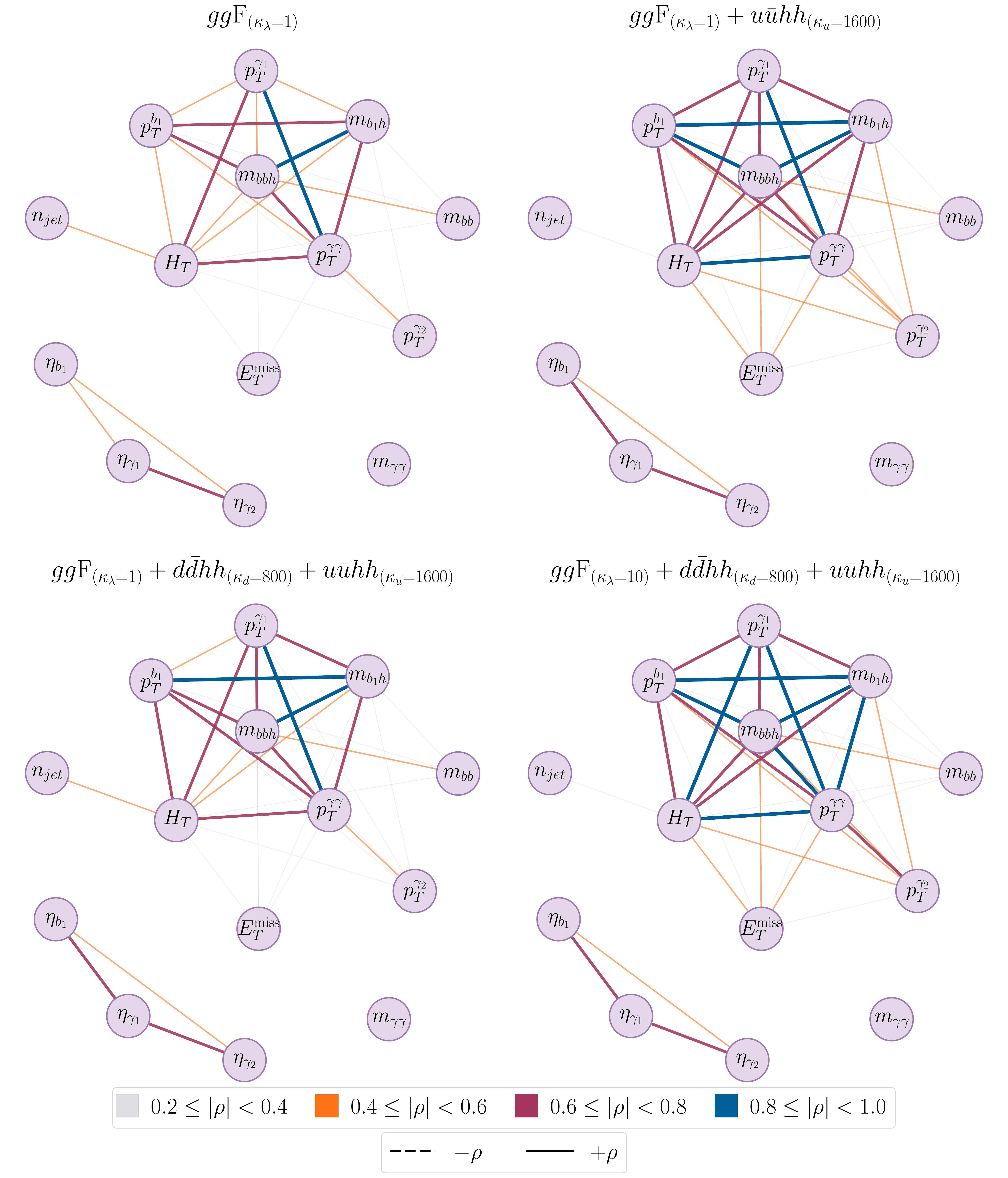}
    \caption{\it Network diagrams visualization of correlations ($\rho$) amongst the kinematic variables used in the analysis. Top left: Only the gluon-gluon fusion channel. Top right: The $gg$F channel along with the $\uuA$ channel with $\kappa_u=1600$. Bottom right: The $\ddA$ channel with $\kappa_d=800$  added to the channels in the top right panel. Bottom left: The same channels as in the bottom right panel but with $\kappa_\lambda=10$.}
	\label{fig:cor-net}
\end{figure}

\noindent This clearly shows that the two channels are distributed quite identically and are difficult to separate. 

Lastly, in the bottom panels of \autoref{fig:shap} we show the variables that are important in separating the $\qqA$ channels from the $gg$F Higgs pair production channel. The invariant mass of the leading $b-$jet and $h$, $m_{b_1h}$ is the most important variable at both HL-LHC and FCC-hh. However, the hierarchy of variables below $m_{b_1h}$ is quite different for HL-LHC and FCC-hh. Both $H_T$ and $p_T^{\gamma\gamma}$ are far less important at FCC-hh than at HL-LHC. This displays the clear advantage that machine learning algorithms have over a cut-and-count analysis where separate cut strategies would have to be built for the two colliders leading to two separate analyses that can, instead, be done with the same framework when using machine learning.

The correlation plots in \autoref{fig:cor-net} show how the linear correlations amongst the variables evolve when different channels are added. In the top left panel are events sampled from the $gg$F distribution. One can already see clustering in some of the variables related to momenta and invariant mass. The other cluster is of the pseudorapidity of the particles in the final state. This correlation structure evolves when one adds the $\uuA$ channel when $E_T^{\rm miss}$ gets connected to the upper cluster in the top right panel. The correlation is now stronger between $\eta_{\gamma_1}$ and $\eta_{b_1}$ and several correlations in the upper cluster are much stronger too. The change in the correlations continues as one keeps adding channels as can be seen from the bottom right and bottom left panels. It is the capture of this change in the correlations (and higher-order correlations) that enhances the capabilities of the machine learning algorithms to distinguish between the various channels. While $m_{\gamma\gamma}$, by its shape alone, allows for the separation between $\bbaa$ and the other channels, the correlations between the other kinematic variables aid in the separation of the channels with one or two Higgs in the final state. 

\subsection{Additional constraints on light-quark Yukawa couplings}
\label{sec:addconst}

There are additional proposed measurements of the light-quark Yukawa couplings that might become relevant at HL-LHC or FCC-hh, a careful study of which is beyond the scope of the current work. We will attempt to include a discussion here so as to provide a comparison with our study and to put it into proper context. 

Studies of rare Higgs decays, involving radiative decays to quarkonia have been proposed in~\cite{Bodwin:2013gca,Kagan:2014ila,Konig:2015qat}, as a direct probe for light Yukawa couplings. These studies were followed upon with experimental searches for such decays~\cite{CMS:2018gcm} and set bounds on the branching ratios, $\mathcal{B} (h \to J/\Psi, \gamma/Z) \sim 10^{-4} - 10^{-6}$ at 95\% CL. More recent bounds on $\kappa_c$~\cite{CMS:2019hve,ATLAS:2022ers} sets it to about $|\kappa_c| < 8.5$ which is hitherto the most stringent \emph{direct} bound on charm quark Yukawa coupling.  Another probe for light-quark Yukawa couplings is the associated production of Higgs with a jet. This channel has been shown to be sensitive to $\kappa_c\sim 1$ when using charm-tagged jets~\cite{Brivio:2015fxa}. Moreover, by looking at differential distributions for this channel, it is possible to obtain stringent bounds on the first-generation Yukawa couplings~\cite{Soreq:2016rae,Bishara:2016jga, Bonner:2016sdg}. Limits on light-quark Yukawa couplings can also be extracted by studying the untagged branching ratios of the Higgs decay to di-jets, under the assumption that no additional new physics present~\cite{Carpenter:2016mwd}. However, all the channels mentioned before, suffer from degeneracy amongst up- and down-quark Yukawa couplings. Other channels can be considered as complementary to them to break this degeneracy. It was shown in Ref.~\cite{Yu:2016rvv,Yu:2017vul}, that the charge asymmetry of the process $pp \to h W^+$ vs. $ pp \to h W^-$ can be used as a probe for light-quark Yukawa couplings as well as to break the degeneracy amongst quark flavours. Moreover, the rare process $pp \to h \gamma$ is also a possible way to distinguish between enhancements of the up- and down-quark Yukawa couplings~\cite{Aguilar-Saavedra:2020rgo}. A cut-based analysis of Higgs pair production using the same final state considered in this work, $hh\to b \bar b \gamma \gamma$, has obtained constraints on light-quark Yukawa couplings~\cite{Alasfar:2019pmn}. The analysis can be thought of as being complementary to the previously mentioned ones as it mostly probes the coupling between two Higgs bosons to quarks. In  addition, we see in  our work that using machine learning significantly improves upon the cut-based analysis. Hence this analysis, taken together with other proposals, provides a probe of non-linearities between the Higgs and light quarks parameterized by the electroweak chiral Lagrangian. Constraints for Higgs couplings could also come from processes that do not involve Higgs production directly. Three-boson production $VVV$ has been shown to give strong projected bounds for light-quark Yukawa couplings for HL-LHC, with ten-fold improvement expected at FCC-hh~\cite{Falkowski:2020znk,Vignaroli:2022fqh}.

\begin{figure}[t!]
	\centering
	\includegraphics[width=\linewidth]{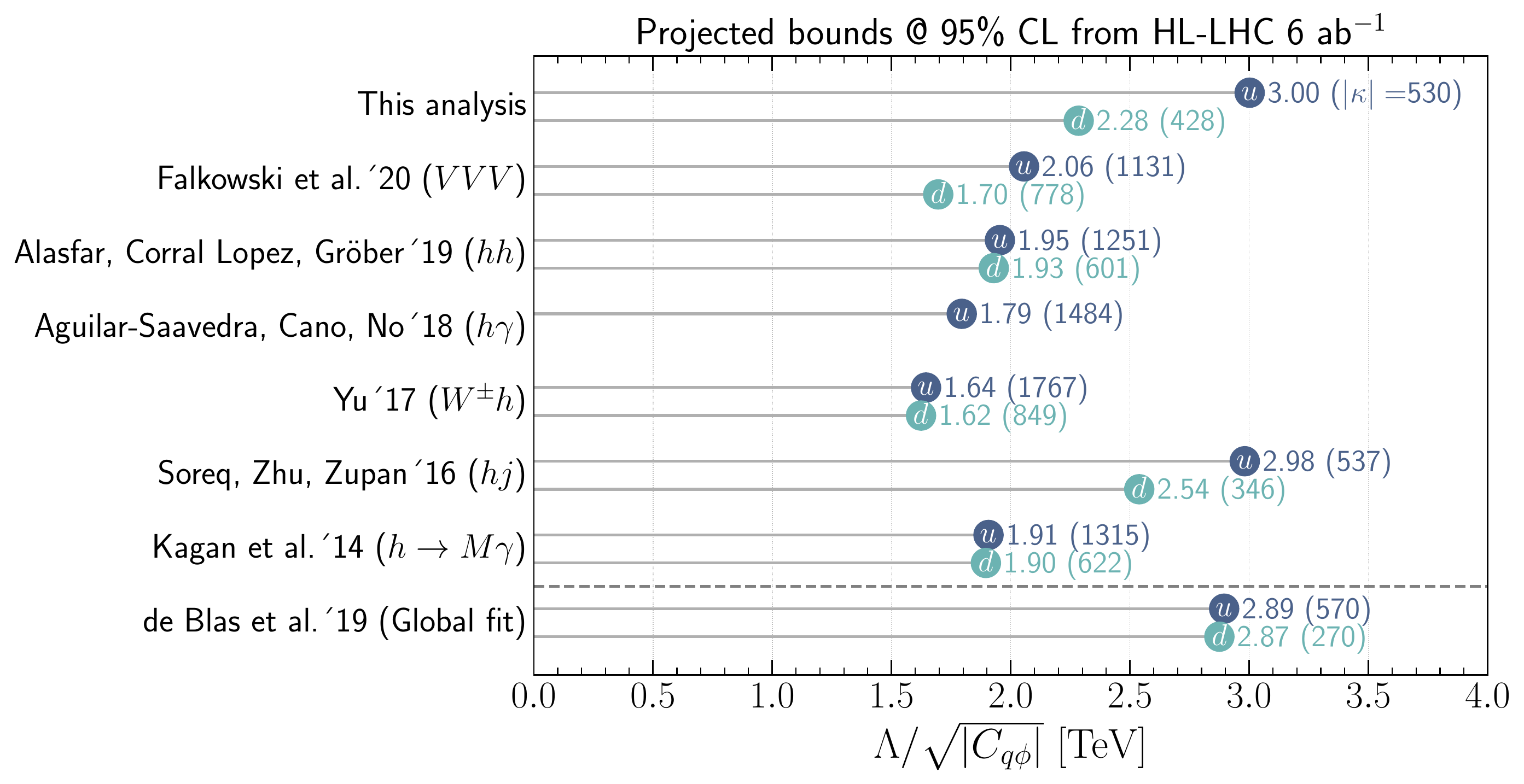}
	\caption{\it Summary of the $95\%$ CL sensitivity bounds on the SMEFT Wilson coefficients $C_{u\phi}$ (blue), and $C_{d\phi}$ (green). The bounds are interpreted in terms of the NP scale $\Lambda$ that can be reached through the measurements of the Wilson coefficient at the HL-LHC at $6 \inab$, the corresponding $\kappa_q$'s are shown inside the parentheses. The $95\%$ CL bounds from single parameter fits are used from this analysis for comparison with previous studies.}
	\label{fig:comparison}
\end{figure}

We present a numerical comparison of the strongest bounds from HL-LHC on the first-generation Yukawa couplings from the studies discussed above in  \autoref{fig:comparison} and compare them to the global fit bounds that have been obtained with no invisible or untagged Higgs
decays allowed~\cite{deBlas:2019rxi}. For $C_{d\phi}$, the most stringent bound comes from the global fit, and the $h+j$ channel, as a model-independent bound, while our analysis provides the second most stringent model-independent bound. For $C_{u\phi}$, our analysis provides the most stringent constraint while the bound from $h+j$ and the global analysis are comparable. The figure is interpreted in terms of the reach of NP scale $\Lambda$ that can be achieved by the measurement of these Wilson coefficients. For future colliders, like the FCC-hh at $100$ TeV, in addition to Higgs pair production, triple Higgs production might be an interesting channel for constraining the operators with Wilson coefficient $C_{u\phi}$ and $C_{d\phi}$ due to the energy increase of a Feynman diagram coupling the quarks to three Higgs bosons. Finally, we note that there are also signatures from experiments not based on colliders for enhanced light-quark Yukawa couplings manifesting in frequency shifts in atomic clocks from Higgs forces at the atomic level~\cite{Delaunay:2016brc}. 
\section{Summary}
\label{sec:Sum}

In this work, we walk through an analysis of how kinematic shapes can be used to glean information about the nuances of various production modes with the same final states but deformed differentially by the existence of degrees of freedom beyond the Standard Model. We show that this information can be extracted by using an interpretable machine learning framework which is not only very effective in separating these differences in kinematic shapes but also yields itself to interpretations in terms of physics that are known and well understood. The example we chose is Higgs pair production in the $b\bar{b}\gamma \gamma$ final state.

We emphasized that probing Higgs pair production is an important next step for an understanding of the model underlying the fundamental interactions of particles and hence a potential gateway to new physics. We show that even beyond the trilinear Higgs couplings, the light-quark Yukawa couplings can be probed through this production mode. In fact, the $\qqA$ channel opens up only in the presence of BSM physics and well-motivated models of new dynamics, bringing about the simultaneous modification of the trilinear Higgs coupling and the light-quark Yukawa couplings. Indeed, we motivated our study by showing that in different frameworks large modifications of the light-quark Yukawa couplings can be obtained. Knowing the difficulty of measuring these couplings, we propose an interpretable machine learning framework that significantly outperforms traditional cut-based analyses.

As opposed to using black-box models, the interpretable framework allows us to gain physics insights into how signal and background separation can be brought into effect, pointing to kinematic variables like $H_T$ and $m_{\gamma\gamma}$ as being important variables that instrument this separation. As a result, we find enhanced sensitivities to $C_\phi$ or $\kappa_\lambda$ that quantify the modification to the Higgs trilinear coupling. Furthermore, we see that the measurement of the light-quark Yukawa couplings is aided by using the methods we advocate bringing about greater sensitivities than would be possible with a cut-based analysis at the HL-LHC and the FCC-hh. The advantage of using an interpretable framework using Shapley values is that it provides added confidence to the robustness of the multivariate analyses that we perform using simulated data.

The salient results of this work are:
\begin{itemize}
\itemsep0em
    \item The modification of the Higgs trilinear coupling can be measured at $\mathcal{O}(1)$ precision at the HL-LHC and at $\mathcal{O}(1\%)$ precision at the FCC-hh. 
    \item The rescaling of the light-quark Yukawa couplings, $\kappa_u$ and $\kappa_d$, can be measured to $\mathcal{O}(100)$ at the HL-LHC and $\mathcal{O}(10)$ at FCC-hh.
    \item The measurement of $\kappa_\lambda$ is significantly diluted once the light-quark Yukawa couplings are allowed to vary. Hence, in a joint fit, the bounds on $\kappa_\lambda$ are weaker. 
    \item There are theoretical models that motivate the simultaneous modification of the trilinear Higgs coupling and the light-quark Yukawa couplings. Hence, the dilution of the bounds on $\kappa_\lambda$ due to the presence of new physics in the light-quark Yukawa sector should be taken into consideration in future phenomenological extraction of $\kappa_\lambda$.
    \item The bounds obtained with the interpretable machine learning framework that we use not only outperforms cut-based analyses but also allow for physics insights into kinematic distributions of the various channels that help distinguish them in an experiment.
\end{itemize}

In conclusion, we stress that the interplay between the Yukawa sector and the Higgs trilinear coupling is non-trivial and requires careful consideration. Future experiments at the HL-LHC and FCC-hh will bring significant improvements in the sensitivities to $\kappa_\lambda$, $\kappa_u$ and $\kappa_d$ through the Higgs pair production channel. In particular, the bounds on the light-quark Yukawa couplings from Higgs pair production can possibly be the most stringent bounds amongst all other experimental probes of the light-quark Yukawa couplings as a result of the evolution of the parton luminosity functions between the single Higgs threshold and the characteristic energy scale in Higgs pair production.

\acknowledgments

This work benefited from support by the Deutsche Forschungsgemeinschaft under Germany's Excellence Strategy EXC 2121 ``Quantum Universe" -- 390833306 as well as under the grant 491245950. L.A. 's research is funded by the Deutsche Forschungsgemeinschaft (DFG, German Research Foundation) - Projektnummer 417533893/GRK2575  ``Rethinking Quantum Field Theory". The work of A.P. is funded by Volkswagen Foundation within the initiative ``Corona Crisis and Beyond -- Perspectives for Science, Scholarship and Society'', grant number 99091. Part of the work done by A.P. was funded by the Roux Institute and the Harold Alfond Foundation. R.G. acknowledges support from a departmental research grant under the project ``Machine Learning approach to Effective Field Theories in Higgs Physics''. This research was supported in part through the Maxwell computational resources operated at DESY, Hamburg, Germany.

\appendix

\section{Discussion of theoretical and systematic uncertainties}
\label{sec:errors}

\begin{figure}[t!]
	\centering
	\includegraphics[width=0.4\linewidth]{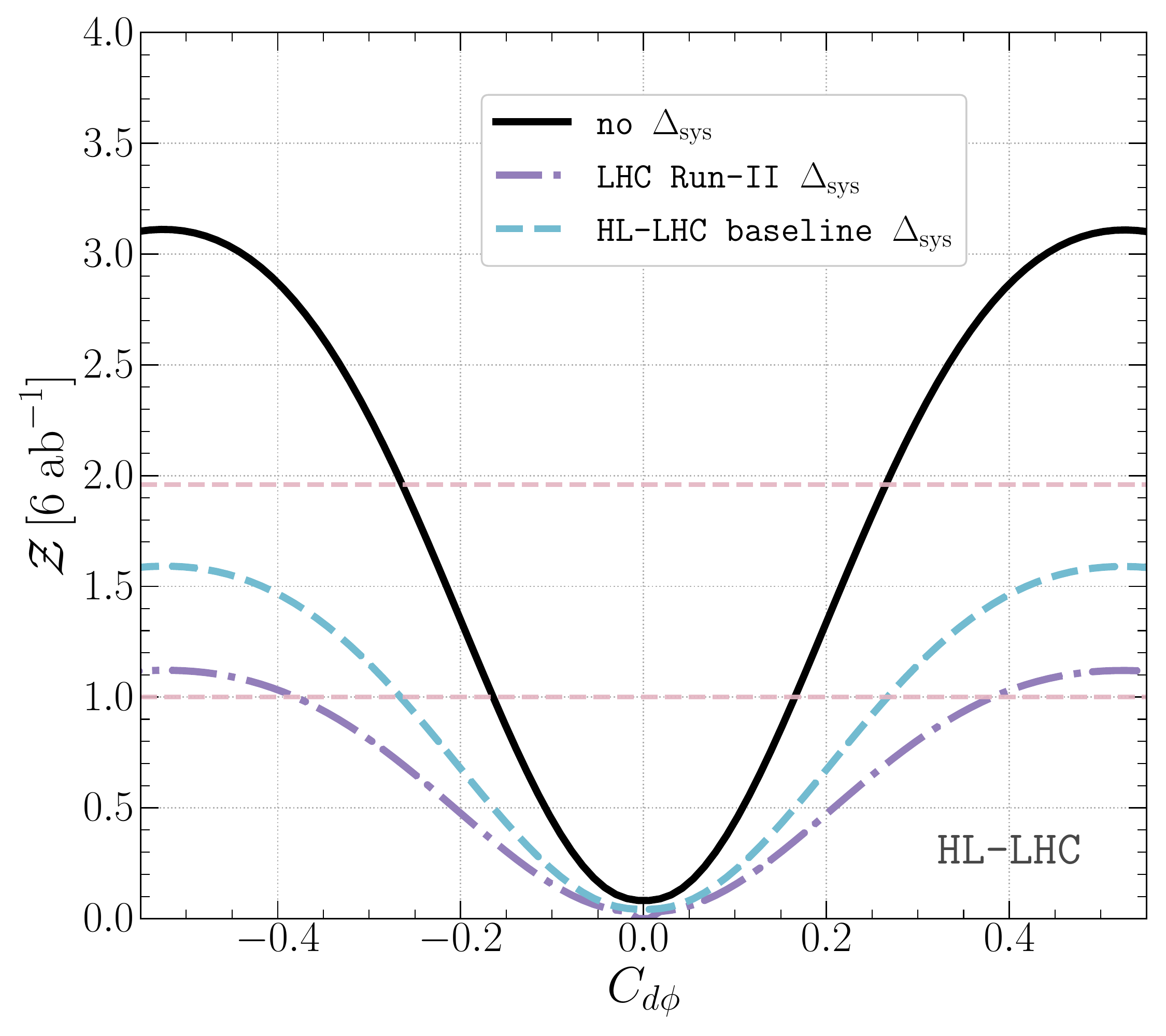}
	\includegraphics[width=0.4\linewidth]{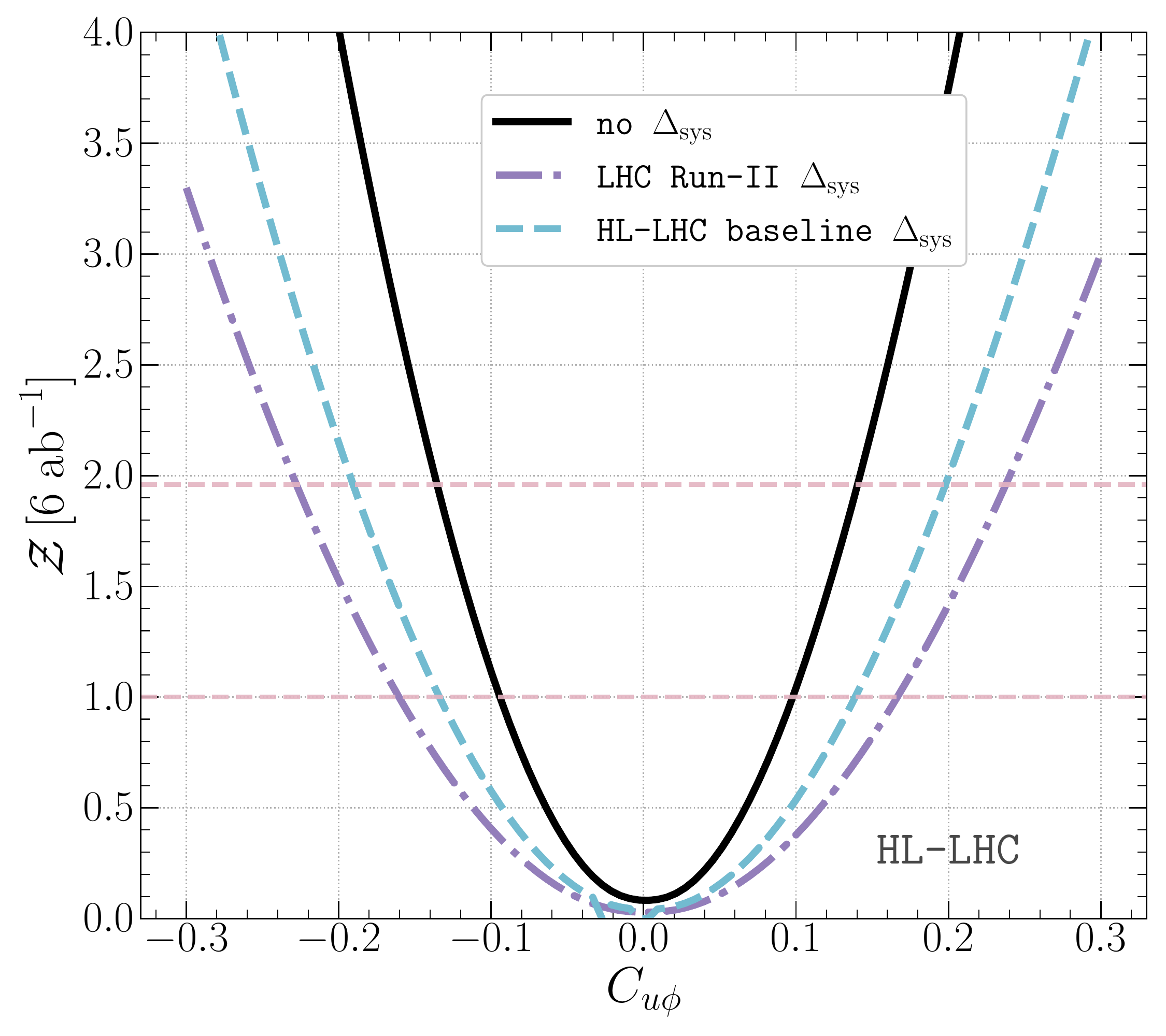}
	\includegraphics[width=0.4\linewidth]{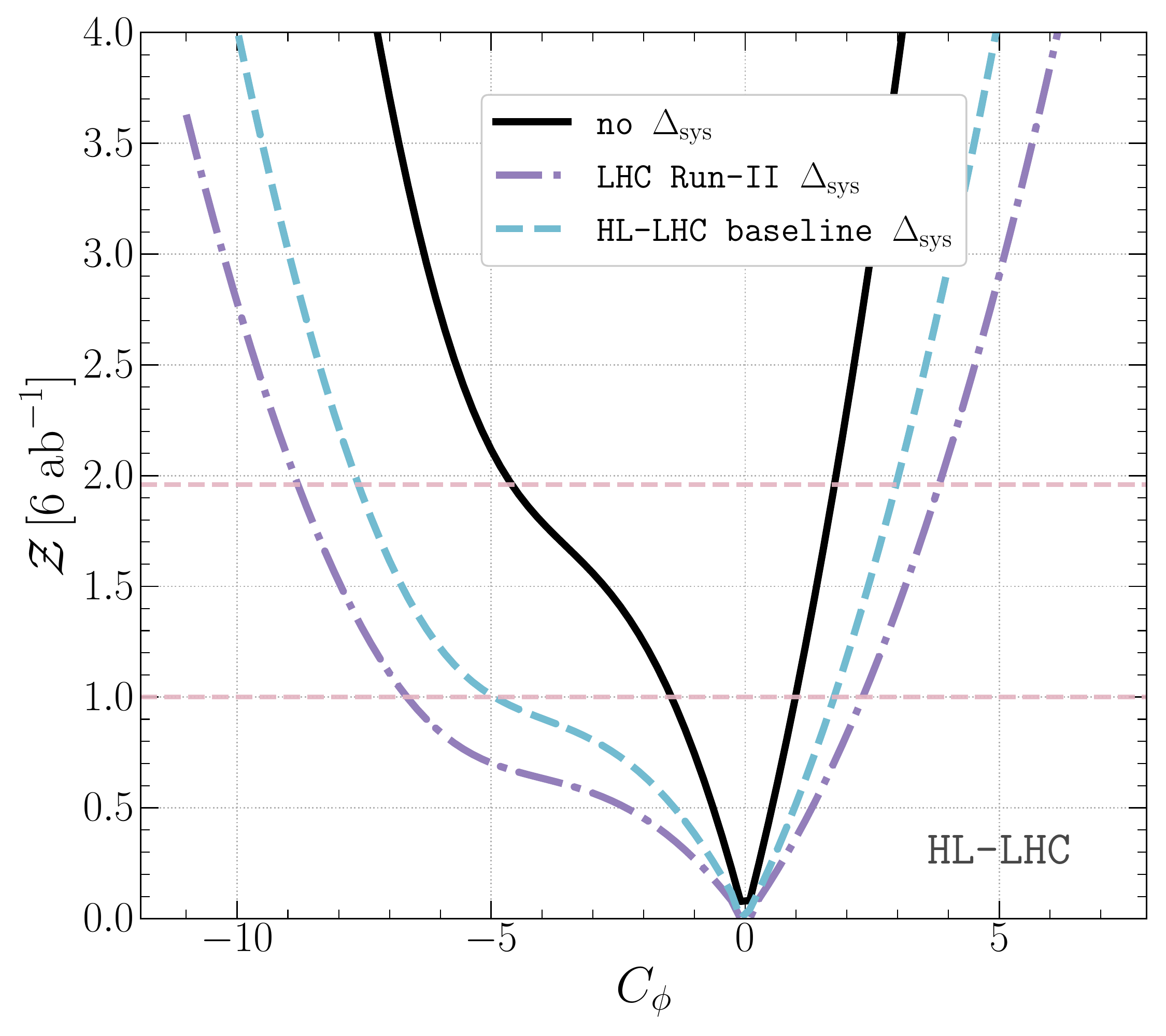}
	\caption{\it The significance, $\mathcal{Z}$, from a single parameter fit for $C_{d\phi}$ (upper left panel), $C_{u\phi}$ (upper right panel) and $C_\phi$ (lower center panel) for the HL-LHC with  no systematic uncertainties (black) and two ansatz for systematic uncertainties. The first is the current Run-II 8.2\% in violet and the HL-LHC baseline 5.3\% estimated by ATLAS in blue, including theoretical uncertainties without the one stemming from the top mass renormalization scheme.}
	\label{fig:systematics}
\end{figure}

In this section, we present an estimate of the systematic uncertainties that can affect the measurements discussed in this work at the HL-LHC. We do not present these estimates for the FCC-hh for lack of sufficient information or the ability to project such uncertainties far into the future. We use two scenarios for systematic uncertainties: the first is a $8.2\%$ uncertainty which corresponds to the current systematic uncertainty that ATLAS has reported for their Run-II search for Higgs pair production~\cite{ATLAS-CONF-2021-016}. The second scenario is the ATLAS HL-LHC baseline systematic uncertainty of $5.3\%$ reported in~\cite{ATL-PHYS-PUB-2018-053}. For LHC run-II, statistical uncertainties remain the dominant part of the uncertainty budget for di-Higgs analysis. Regarding the systematic uncertainties, experimental sources remain the dominant part in comparison to the theoretical ones.  The story flips for the HL-LHC where the main source of uncertainties is expected to be coming from theoretical uncertainties. The current theoretical uncertainty estimate of the SM gluon fusion process at NNLO is ${}^{+6\%}_{-23\%}$ for $\sqrt{s}=14\text{ TeV}$ and ${}^{+4\%}_{-21\%}$ for $\sqrt{s}=100\text{ TeV}$~\cite{Baglio:2020wgt}. The largest part of the uncertainty stems from the uncertainty due to the renormalization scheme choice of the top quark mass. This uncertainty can, for the moment, only be estimated at NLO since no full mass-dependent results at NNLO are available. Moreover, the top quark mass renormalization scheme uncertainty is not included in the estimated HL-LHC (nor LHC Run II) uncertainties schemes that we have considered. 

In \autoref{fig:systematics} we show the significance~$\mathcal{Z}$ for the three Wilson coefficient, $C_\phi$, $C_{u\phi}$ and $C_{d\phi}$, at the HL-LHC from single parameter fits with no systematic uncertainties (black), LHC Run-II (violet) and HL-LHC baseline (blue) systematic uncertainties ansatz. We observe that for the current Run-II ansatz, the bounds for all three Wilson coefficients is diluted by 100\% or more. As for the HL-LHC baseline, the bounds are diluted by $\sim$ 70\%. However, it should be noted, that both systematic uncertainties scenarios are rather conservative. It is likely that the HL-LHC detector upgrade and new theoretical developments in higher-order corrections to the di-Higgs cross-section will reduce the systematic uncertainties from the baseline. 

\section{Light-quark Yukawa and Self Coupling at Future Lepton Colliders}
\label{sec:Lep}

\begin{table}[t!]
    \centering
    {\small
    \begin{tabular}{cccc}
    \toprule
         Collider &  $|\kappa_u|$ &  $|\kappa_d|$  & $\delta \kappa_\lambda$  ($1\sigma$) \\ \midrule
         240GeV 5ab$^{-1}$ (CECP/FCC) & 192~\cite{Gao:2016jcm}  & 90~\cite{Gao:2016jcm}   & 100\% (Indirect~\cite{DiVita:2017vrr})\\
         350 GeV 1.5 ab$^{-1}$ (FCCee) & 310~\cite{deBlas:2019rxi} & 140~\cite{deBlas:2019rxi} &  40\% (Indirect~\cite{DiVita:2017vrr})\\ 
         500 GeV 4 ab$^{-1}$ (ILC) & 330~\cite{deBlas:2019rxi} & 160~\cite{deBlas:2019rxi}&  27\%~\cite{Bambade:2019fyw} \\
         1 TeV 8 ab$^{-1}$ (ILC) & -- & -- &  10\%~\cite{deBlas:2019rxi} \\
         3 TeV 1 ab$^{-1}$ (CLIC) & 430~\cite{deBlas:2019rxi} & 200~\cite{deBlas:2019rxi} &  10\%~\cite{deBlas:2019rxi}\\
         10 TeV 10 ab$^{-1}$ (Muon) & --& --&  3\%~\cite{deBlas:2019rxi}\\ \bottomrule
         \end{tabular}
    \caption{\it Prospective light-quark Yukawa and Higgs self-coupling sensitivities at future lepton colliders. The light-quark Yukawa bounds are 95\% CL, while the self-coupling bounds are $1\sigma$ or 68\% CL sensitivity reach.}
    \label{tab:exp_yqyl}
    }
\end{table}

Future high energy lepton colliders~\cite{Charles:2018vfv,Bambade:2019fyw,CEPCStudyGroup:2018ghi,Vasquez:2019muw} offer further alternative and clean signals for measurement of Higgs properties. For example, Higgs decays to ``un-tagged'' light jets including $u,d,s$ quarks can be further disentangled from $h\to gg$ using event shape analysis~\cite{Gao:2016jcm} and can reach a sensitivity of $\kappa_d \approx 90$ and $\kappa_u \approx192$ at 250 GeV with 5~ab$^{-1}$ data compared with a sensitivity of $\kappa_d \approx 470$ and $\kappa_u \approx900$ at the 6 ab$^{-1}$ HL-LHC~\cite{Carpenter:2016mwd,Soreq:2016rae}.

The sensitivity to Higgs self-coupling comes indirectly for center of mass energy below 250 GeV from the precision measurement of the $Zh$ production channel ($1\sigma$ bound on $\delta \kappa_\lambda$ of $0.4$ at 250 GeV), and at 500 GeV directly from the $Zhh$ channel ($1\sigma$ bound on $\delta \kappa_\lambda$  of $0.27$ at 500 GeV), and from vector boson fusion like production to $hh\nu\nu$ when 1 TeV or higher energy scales are available ($1\sigma$ bound on $\delta \kappa_\lambda$ of 10\% at 1 TeV). The prospective sensitivity depends on the collider setup, mainly the integrated luminosity and polarization of initial lepton beams. Given the updated prospects of future machine designs~\cite{deBlas:2019rxi}, we list a short summary in \autoref{tab:exp_yqyl} of the expected sensitivities on the individual parameters in the $\kappa$ framework. These numbers are all assuming one-parameter fits in $\kappa$. No simultaneous fit including both $\kappa_q$ and $\kappa_\lambda$ (or using the corresponding SMEFT operators) has been performed yet.

\section{Shapley values}
\label{sec:shapley}

Shapley values~\cite{shapley1951notes} are defined for a game $(v, N)$, where $N=\{1,\ldots,n\}$ is a set of players in the game and $v$ is the characteristic function that assigns a non-negative real value $v(S)$ to every coalition $S \subseteq N$, and zero to the empty coalition, i.e. $v(\emptyset) = 0$. The function $v$ fully describes the game, as it maps players to payoffs. A subset $S$ of $N$ is referred to as a coalition, and $v(S)$ the value of the coalition. The marginal contribution of a player $i$ to the coalition $S$ is defined as $v(S \cup \{i\}) - v(S)$.  The average marginal contribution of player $i$, over the set ${S}_k$ of all coalitions containing of $k$ players which exclude $i$, is
\begin{equation} 
    \overline{v}_k(i) = \frac{1}{|{S}_k|}\sum_{S \in S_k} [v(S \cup \{i\}) - v(S)] \,,
    \label{eq:marginal}
\end{equation}
where $|{S}_k| = \binom{n-1}{k}$. The Shapley value of player $i$ is then
\begin{equation} \label{eq:shap2}
    \phi_i(v) = \frac{1}{n}\sum_{k = 0}^{n - 1} \overline{v}_k(i) \,.
\end{equation}
Combining \autoref{eq:marginal} and \autoref{eq:shap2} we get:
\begin{equation} \label{eq:shap3}
    \phi_i(v) = \sum_{S\subseteq N \backslash \{i\} }\frac{|S|! (n-|S| -1)!}{n!}\left(v(S
    \cup \{i\}) - v(S)\right), \quad i = 1,\dots, n \,;
\end{equation}
a weighted mean over all subsets $S$ (including the empty set $S=\emptyset$) not containing $i$ with $|S|$ denoting the cardinality of $S$. The decomposition into Shapley values is the only solution  satisfying a set of four favorable axioms~\cite{Young1985}: \textit{Efficiency}, \textit{Symmetry}, \textit{Linearity}, and \textit{Null Player} which are given by:
\begin{itemize}
    \item   \textit{Efficiency}: The sum of the payoff to the individual players equal the payoff of the grand coalition, i.e.
    $$\sum_{i\in N} \phi_i(v) = v(N).$$
    \item   \textit{Symmetry}: The contributions of two players $j$ and $k$ should be the same if they contribute equally to all possible coalitions,
    $$v(S\cup\{j\}) = v(S\cup\{k\}) \quad\forall S\subseteq \{1,\ldots,n\}\backslash \{j,k\} \iff \phi_j=\phi_k.$$
    
    \item   The \textit{Null Player}: A player $j$ that does not change the payoff –- regardless of which coalition of players they are added to -- should have a Shapley value of 0,
    $$C(v\cup\{i\}) = v(S) \quad\forall S\subseteq \{1,\ldots,n\} \iff \phi_i=0$$
    \item   \textit{Linearity}: The payoffs for the linear sum of two games is the linear sum of the payoffs for each game. For two games $(v, N)$ and $(w,N)$:
    $$\phi_i(\alpha_vv+\alpha_ww)=\alpha_v\phi_i(v)+\alpha_w\phi_i(w).$$
\end{itemize}

For every coalition that can be formed, value of player $i$ is assessed for when the player is added to the coalition. The averaged payoff that player $i$ gets from all possible coalition is the Shapley value of the player. In a multivariate analysis, the players can be replaced by independent variables and the payoff can be replaced by an outcome or dependent variable(s). Shapley values for the variables can be computed by fitting a machine learning model to data and using the model as the characteristic function. There are distinct challenges to doing this. Firstly, an exact computation of Shapley values scale exponentially with the number of variables and becomes intractable even for a moderate number of variables. Secondly, retraining a model after removing variables is not possible since that would change the model itself and, hence, change the characteristic function. 

As a solution to to the first problem, SHAP~\cite{NIPS2017_7062} uses either kernel methods or tree-explainers~\cite{2018arXiv180203888L,Lundberg:2020vt} which significantly reduces the computational burden. The tree-explainer addresses the second problem by ignoring the branches of a decision tree that contain the variable not included in a coalition and computing the weighted average of the outcome from the rest of the model. Due to the linearity property of the Shapley value, Shapley values from an ensemble of trees can be added to compute the final Shapley value of a variable. Given that the variable which, on an average, consistently contributed more to the outcome will have a higher Shapley value, an importance ranking can be based on this Shapley value which is known as feature importance in machine learning.

\bibliographystyle{JHEP-CONF}
\bibliography{hh}

\end{document}